\documentclass[journal,twocolumn,10pt,twoside]{IEEEtran}

\usepackage{cite}

\ifCLASSINFOpdf
\else
\fi

\usepackage{url}
\usepackage{amsmath}
\usepackage{amssymb}
\usepackage{cite}
\usepackage{graphicx}
\usepackage{algorithm}
\usepackage{algpseudocode}
\usepackage{subfigure}
\usepackage{multirow} 
\usepackage{longtable}
\usepackage{threeparttable}
\usepackage{caption3}
\usepackage{array}
\usepackage{cases}
\usepackage{color}
\usepackage[T1]{fontenc}
\usepackage[utf8]{inputenc}
\usepackage{authblk}
\usepackage{subfigmat}

\usepackage{caption}
\usepackage{multirow}

\usepackage{float}

\usepackage{multirow, makecell}
\usepackage{booktabs}
\usepackage{epstopdf}
\usepackage{graphicx, subfigure}

\begin{document}
%
\title{Wireless Information and Power Transfer for IoT: Pulse Position Modulation, Integrated Receiver, \\ and Experimental Validation}
\author{Junghoon~Kim,~\IEEEmembership{Member,~IEEE,}
    and~Bruno~Clerckx,~\IEEEmembership{Senior~Member,~IEEE}
\thanks{J. Kim and B. Clercks are with the Department
of Electrical and Electronic Engineering, Imperial College London, London,
SW7 2AZ, UK e-mail: (junghoon.kim15,b.clerckx@imperial.ac.uk).

This work was supported in part by the EPSRC of U.K. under Grant EP/P003885/1 and EP/R511547/1.
}
}

%

\maketitle

\begin{abstract}
Simultaneous wireless information and power transfer (SWIPT) has emerged as a viable technique to energize and connect low-power autonomous devices and enable future Internet of Things (IoT). 
A major challenge of SWIPT is the energy consumption of the receiver of such low-power devices. 
An attractive low-power solution consists in an integrated information decoder (ID) and energy harvester (EH) architecture for SWIPT receiver (IntRx) where the received RF signal is first rectified before being used for information decoding. 
Such architecture eliminates the need for energy-consuming RF components such as local oscillators and mixers.
This paper proposes a novel modulation and demodulation method for the IntRx SWIPT architecture based on pulse position modulation (PPM) where information is encoded in the position of the pulse.
The new method transmits high amplitude pulses to increase the Peak-to-Average Power Ratio (PAPR) of the transmit signal and exploits the EH's nonlinearity so as to boost the harvested DC power.
Simultaneously, the information can be decoded from the rectifier signal by simply finding the position of the pulse in a certain symbol duration. 
We have analyzed both the information and the power transfer performance of the newly proposed PPM for IntRx SWIPT theoretically, numerically, and experimentally.
To that end, we have established a SWIPT system testbed in an indoor environment by prototyping a base station to transfer information-power signal and the IntRx SWIPT receiver including ID and EH blocks.
The performance evaluation of the PPM was carried out in various conditions, and the results have been compared and contrasted to conventional signals.
Theoretical, numerical, and experimental results highlight the significant benefits of the proposed PPM scheme to enhance the power transfer performance and operate information decoding with low-power consumption.
\end{abstract}


%
\IEEEpeerreviewmaketitle

\section{Introduction}

\IEEEPARstart{W}{ireless} information and power transfer (WIPT) is an emerging paradigm to connect and energize devices wirelessly in the future Internet of Things (IoT) and wireless sensor networks (WSN) \cite{Clerckx2019}.
Radiofrequency (RF) waves have traditionally been used for wireless information transfer (WIT) or wireless communications.
However, RF waves can carry both energy and information. 
Recently, thanks to the reduction in power consumption of electronics and the need to energize a massive number of low-power devices, far-field RF wireless power transfer (WPT) has attracted attention as a feasible and promising power supply technology \cite{Koomey2011}.
RF WPT is an advantageous power source for low-power wireless devices since it is controllable, enables long-distance power delivery, and can be implemented in a small form factor at the receiver compared to other technologies such as inductive coupling or magnetic resonance.
In addition, WPT can be integrated with WIT so as to exploit RF waves for the dual purpose of communicating and energizing.
\par
Simultaneous information and power transfer (SWIPT) is one of the subset categories of WIPT and refers to the simultaneous transmission of information and power from the transmitter to the receiver.
It is suitable for a system that is comprised of a base station serving low-power wireless devices or sensors, such as in IoT networks. 
The base station is able to transmit information signals while simultaneously energizing low-power wireless devices. 
SWIPT was first studied in \cite{Varshney2008}, which showed that there exists a fundamental trade-off between information rate and delivered energy, the so-called rate-energy (R-E) region.
The R-E region is a function of many factors, including the signal design, the wireless channels, the transmitter and receiver architecture, the energy harvester architecture and modeling, and the network architecture \cite{Clerckx2019}.
Numerous studies have been conducted in various disciplines to improve the SWIPT system performance in terms of both information transfer rate and power transfer efficiency.
\par
WPT is an important building block of SWIPT.
The vast majority of WPT research in the RF literature has been devoted to the design of efficient energy harvesters (e.g., rectennas combining rectifiers and antennas) so as to increase the RF-to-DC conversion efficiency \cite{Georgiadis2010,Sun2012,Valenta2014,Lee2015}. 
Early RF experiments also showed that the harvested energy depends not only on the rectenna design but also on the input signal to the rectenna \cite{Trotter2009,Boaventura2011,Collado2014}. 
Consequently, WPT signal design strategies have emerged in the communication and signal processing literature since techniques like beamforming, waveform, modulation, channel estimation heavily influence the energy harvesting performance \cite{Clerckx2021futureWPT,Zeng2017}.
In \cite{Clerckx2016}, novel channel-adaptive beamforming and waveform design strategies have been proposed using an analytical model of the diode rectifiers to exploit a beamforming gain, the frequency selectivity of the channel, and the rectifier nonlinearity.
Energy modulation strategy for WPT has been studied in \cite{Varasteh2020} where Gaussian distributed signal and on-off keying (OOK) signaling (with low-probability of a high amplitude signal) were shown to exploit the rectifier nonlinearity and consequently enhance the power conversion efficiency at the energy receiver.
Transmit diversity strategy, which induces artificial fluctuation of the wireless channel using multiple antennas, has been proposed in \cite{Clerckx2018} and shown to improve the RF-to-DC conversion efficiency thanks again to the rectifier nonlinearity.
Those WPT signal design strategies were validated experimentally in a real-world wireless environment in \cite{Kim2020}, and the experiment also confirmed that the rectifier nonlinearity is beneficial to the performance in the low-power regime and is essential to accurately design, analyze and predict the WPT system performance.
In particular, the experimental results in \cite{Kim2020} clearly showed that signals with a high peak-to-average power ratio (PAPR) stimulate the nonlinear characteristic of a diode and improve the output DC power. 
For example, some signal design strategies such as OOK signaling and the transmit diversity are all techniques that increase the PAPR of the signal incident to the WPT receiver.
\par
Earlier works on SWIPT have characterized under simplifying energy harvester assumptions the R-E region in MIMO broadcasting channels \cite{Zhang2013, Xu2013} and broadband systems \cite{Huang2013, Ng2013}. 
The system architectures and signal distribution schemes of SWIPT receivers, which include an information decoder (ID) and energy harvester (EH), have been shown to impact the R-E trade-off and system performance. 
Time-switching (TS) and power-splitting (PS) schemes are common receiver architectures widely used in the SWIPT literature \cite{Clerckx2019} because they can be simply implemented by connecting conventional energy and information receiver modules to RF switch or power splitter.
However, significant R-E trade-offs are inevitable in those schemes because the received RF signal is not entirely delivered to both the EH and ID; each part only receives a certain portion of the RF signal in the form of time or power split.
In order to maximize the harvested DC power and data rate while using the TS and PS architectures, a new line of research on signal design for SWIPT has emerged \cite{Clerckx2019}.
One of the approaches is to design a dedicated modulation scheme for SWIPT with TS and PS receivers, and several techniques have been proposed and evaluated by simulations in \cite{Zhu2017, Kim2018a, Bayguzina2019}.
Another approach to enlarge the R-E region is to use superimposed signaling, which overlaps WPT and WIT signal with a specific amplitude ratio obtained through optimization \cite{Clerckx2018a, Kang2018, Kang2019}.
The improved system performance and R-E region expansion using the superimposed SWIPT signaling have been verified experimentally as well as simulations in \cite{Kim2019}.
\par
Though less investigated in the communication literature, a different type of receiver architecture with integrated EH and ID receiver (IntRx) has also been proposed \cite{Zhou2013}. 
In the IntRx architecture, the received RF signal is first converted and rectified into DC voltage signal, and then a portion of this rectified signal is used for information decoding. 
In contrast to the TS and PS receiver architectures, the received RF signal of the IntRx architecture is not divided and distributed to the EH and ID blocks but is fully used and converted from the RF to DC power.
This behavior is beneficial to improve the power delivery performance of the SWIPT system, but the information performance is affected. 
Indeed, the rectifying process from RF to DC is similar to an envelope detection process and can be regarded as equivalent to the down-conversion from RF to baseband in the conventional information receiver.
The key difference is that the rectified voltage signal has to be much smoother than the envelope signal to serve as a DC power source in the SWIPT receiver. 
Hence, the rectified signal loses most of its original characteristics, and it causes a degradation of the information decoding performance compared to conventional information receivers. 
However, a considerable benefit of the IntRx receiver is that it can significantly reduce the power consumption of low-power SWIPT receivers because the IntRx receiver does not require energy-consuming RF components such as local oscillators and mixers which are needed to downconvert RF to baseband signals in conventional RF receiver.
This is a major advantage for the widespread adoption of SWIPT in low-power nodes. 
To take advantage of these benefits in SWIPT, several modulation schemes suitable for the IntRx structure have been proposed. 
A technique that transmits a high PAPR signal to increase the DC output of the EH and decodes the information by detecting the PAPR of the received signal was proposed in \cite{Kim2016}.
Various schemes for decoding information by detecting several different factors from the signal rectified at the receiver such as the amplitude \cite{Claessens2018a}, the intermodulation of multi-frequency transmission signal \cite{Rajabi2018,Claessens2019}, and the number of tones \cite{Krikidis2019} were proposed, and their performance was evaluated.
The power delivery performance enhancement of the schemes proposed so far is still limited because the schemes change the signal's PAPR, amplitude, or the number of tones to convey information, which hinders maintaining the shape of the signal that maximizes the power transfer efficiency. 
Moreover, some schemes are only operative under a certain channel condition, such as a frequency-flat channel environment. 
\par
In this paper, we propose a novel signal design method for SWIPT with IntRx architecture that is driven by the recent literature on signal design for WPT so as to maximize the output DC power at the receiver while enabling information decoding without power-consuming RF components.
In WPT, OOK signaling was shown to increase the harvested DC power significantly \cite{Varasteh2020, Kim2020}.
The OOK signaling method transmits with a low probability large amplitude signals for short periods of time, and during the remaining time, it does not transmit any signals.
Consequently, the PAPR of the signal is large and leads to an increase in the harvested DC power at the receiver.
There exists a modulation scheme that has a nearly identical waveform shape as the OOK signal among the conventional WIT signals, which is pulse position modulation (PPM) that uses the time shift of a pulse to distinguish information.
The PPM is one of the pulse modulation techniques widely used for information transfer, which transmits a pulse for only a certain chip of time corresponding to the encoded information during the entire symbol duration \cite{Carbonelli2006, Garrett1983}.
In other words, the information is encoded on the pulse position during the symbol period of the time domain signal, and all the transmit signal power is concentrated on the corresponding pulse, thus increasing the PAPR of the signal and enabling to improve harvested DC power at the receiver.
In terms of information transfer, the PPM is a non-coherent modulation that does not need a power-consuming local-oscillator at the receiver; thus, it is suitable for low-power IntRx SWIPT receiver.
From those observations, we have devised a novel SWIPT modulation and signal design scheme called M-PPM for SWIPT, where M is a modulation order, that is inspired by PPM and OOK signaling from conventional WIT and WPT, respectively. 
The new method was designed by modifying each method of PPM and OOK to adapt to the SWIPT system and requirements, as well as to maximize the efficiency of the information and power transfer.
In the rest of this paper, we provide principles of the newly proposed M-PPM for SWIPT and evaluate its performance in terms of information and power transfer. 
The feasibility of the new modulation in the practical environment is also validated using realistic simulation and experiment. 
The contributions of this paper are summarized as follows.
\par
\textit{First}, we propose a new modulation and demodulation technique, called M-PPM, for SWIPT to enhance the information and power transfer performance of the IntRx SWIPT architecture.
M-PPM for SWIPT uniquely leverages the benefits of OOK signaling for WPT and conventional PPM for WIT. 
In order to increase the DC power harvested at the low-power SWIPT receiver, the waveform shape of the novel modulation signal is designed to have a high PAPR based on the waveform shape of conventional PPM and OOK. 
This high PAPR leads to high ripple voltage during the rectification process, so we have devised an information decoding method based on the PPM that decodes information leveraging the peak position of the ripple voltage.
The proposed method allows the receiver to decode information without power-consuming RF components and is suitable for low-power IntRx SWIPT systems.
Uniquely, the newly proposed M-PPM is a practical modulation and demodulation technique with a finite-constellation unlike the conventional OOK signaling, and is capable of decoding information from a rectified DC voltage signal unlike the existing PPM for WIT.
In contrast to other modulation techniques proposed for the IntRx SWIPT system such as \cite{Kim2016}, the PAPR of the M-PPM signal is not changed for different information symbols, so the output DC power at the receiver can be maximized in all situations by maintaining a high PAPR waveform shape.
This is the first paper on such modulation for SWIPT and the first paper to recognize the usefulness of PPM for SWIPT.
\par
\textit{Second}, we design and implement an entire SWIPT system prototype capable of generating modulation signal, transmitting over-the-air, harvesting DC power from the received signal, and demodulating information.
This work is the first attempt to implement IntRx SWIPT system operative under a realistic indoor wireless environment and several meters of distance between transmitter and receiver, and also the first implementation of the proposed M-PPM for SWIPT modulation technique on the physical equipment.
We used software-defined radio (SDR) equipment to generate a modulated RF transmission signal and designed an IntRx SWIPT receiver that harvests DC power and decodes the information from the received RF signal at the same time.
The prototype was installed in an indoor laboratory environment, allowing the evaluation of the performance of the newly proposed SWIPT signal and system in a realistic environment.
\par
\textit{Third}, the performance of the proposed M-PPM for SWIPT modulation is analyzed and confirmed using realistic circuit simulations and experimentation. 
WPT and WIT performance of the new M-PPM modulation scheme, as well as the trade-off between them, are thoroughly evaluated under realistic conditions.
The simulation was carried out using joint MATLAB and Pspice software to simulate both the effects of wireless channels and rectifier circuits under various conditions.
The experiments were conducted using both a cable connection to verify the information and power transfer performance according to the RF power incident to the receiver by sweeping the power setting and an over-the-air environment to confirm the proper operation and performance evaluation in the realistic wireless channel.
Experimental results in a real-world condition confirm the feasibility and outstanding information and power transfer performance of the new modulation technique.
\par
The rest of this paper is organized as follows. 
Section \ref{sec:sys_model} presents the system and signal model of the SWIPT system with integrated receiver architecture. 
In Section \ref{sec:modulation}, we describe the principle of the proposed MPPM modulation scheme and provide examples of the actual signal generation and information decoding from the signal.
Performance evaluations of the modulation scheme through circuit simulation and experiment are presented in Section \ref{sec:simulation} and \ref{sec:experiment}, respectively. 
Then Section \ref{sec:conclusion} concludes the work.
%

\section{SWIPT with Integrated Receiver System Model}\label{sec:sys_model}
\begin{figure}[t]
	\centering
	\includegraphics[width=0.47\textwidth]{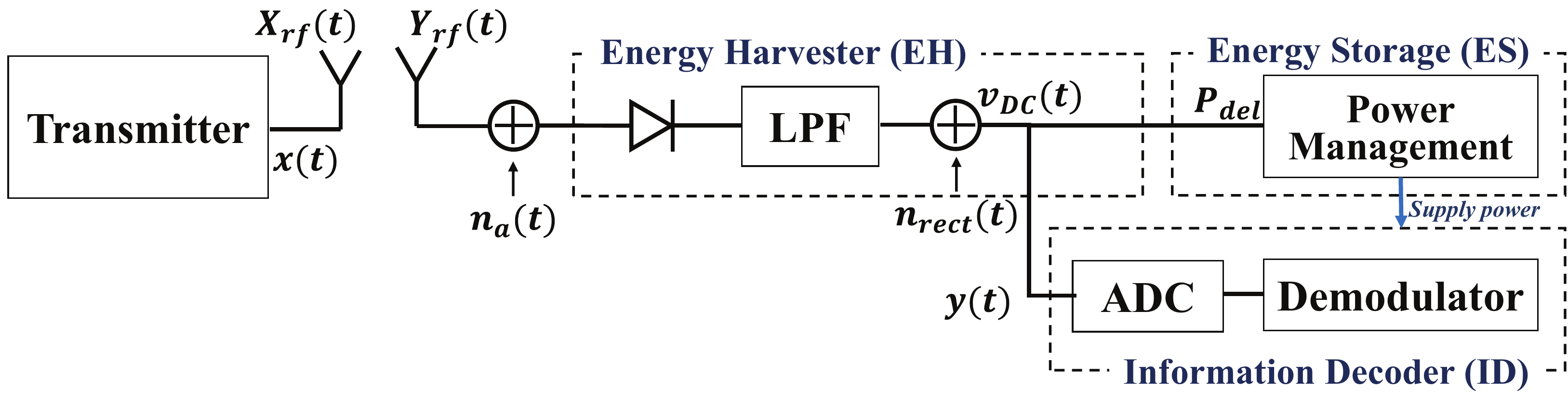}
	\caption{System architecture of SWIPT with EH and ID integrated receiver (IntRx).}
	\label{fig_systemmodel}
\end{figure}

In this section, we introduce a SWIPT architecture with IntRx receiver and briefly describe the functioning of the various building blocks of the system. 
We consider a single-user point-to-point SISO SWIPT system in a general multipath environment.
The overall system architecture of the WIPT with IntRx receiver is presented in Fig. \ref{fig_systemmodel}. 
\par
Both transmitter and receiver are equipped with a single antenna to wirelessly convey information and energy simultaneously. 
The transmitter is capable to modulate, generate, and radiate RF signals like the transmitter used in the conventional WIT system.
The receiver is equipped with an energy harvester and an information decoder and is capable of harvesting the DC power (denoted by $P_{del}$) from the received RF signals as well as decoding the information simultaneously. 
The information decoder consists of an ADC and a digital signal processing unit to digitize the harvested voltage signal and then demodulate conveyed information.
Unlike the PS/TS SWIPT receiver, the IntRx structure allows energy harvesting and information decoding without splitting signals in the time or power domains. 
The RF signal radiated from the antenna of the transmitter is modeled as 
\begin{equation}
X_{\mathrm{RF}}(t) = \sqrt{2}\Re \{x(t)e^{j2\pi f_{c}t} \},
\end{equation}
where $x(t)$ is the baseband information-power signal which is modulated by the PPM scheme and $f_{c}$ is the carrier frequency. 
We assume that the baseband information-power signal $x(t)$ is subject to an average transmit power constraint $P$, i.e., $\mathbb{E}\{\lvert x(t) \rvert ^{2} \} \leq P$. 
More detailed descriptions of the modulation design and signal model of the baseband signal $x(t)$ are provided in the next section.

\par
The RF signal propagates through the wireless channel and is received at the antenna of the receiver. 
The received RF signal $Y_{\mathrm{RF}}(t)$ can be written as 
\begin{equation}
Y_{\mathrm{RF}}(t) = \sqrt{2}\Re \{ \Lambda^{-1/2} h x(t)e^{j2\pi f_{c}t} \},
\end{equation}
where $ \Lambda^{-1/2} h$ denotes the wireless channel between the transmitter and receiver with $ \Lambda $ the path loss and $ h $ the complex fading coefficient.
The $Y_{\mathrm{RF}}(t)$ is first fed into the EH and is converted to the DC voltage signal $v_{DC}(t)$. 
We assume that the antenna noise $n_{a}(t)$ is much smaller than the rectifier noise $n_{rect}(t)$, and it is filtered during rectification. 
Thus, the effect of the antenna noise $n_{a}(t)$ to the rectifier output voltage $v_{DC}(t)$ is negligible. 
The signal input to the information decoder is modeled as 
\begin{equation}
y(t) = v_{DC}(t) + n_{rect}(t),
\end{equation}
which is playing the same role as the baseband information signal in a conventional wireless information receiver.
In this IntRx structure, the baseband signal $y(t)$ is obtained through rectification in the EH without power-consuming RF components such as local oscillator and mixer.
In other words, extracting the baseband information signal from the received RF signal is a power-consuming process in the conventional WIT receiver, but it is transformed into the process of harvesting power in the IntRx structure. 
In addition, DC power and baseband information signals can be obtained from a single source of RF signal, unlike the power and time splitting used in the TS or PS receiver structure.
Also, since the input impedance of the ADC's input port is sufficiently large, it is assumed that the power leaked to the input port of the information decoder (ID) is very small and can be neglected.
Many commercial ADC chip specifications state that its input leakage current levels are less than several $\mu $A, so the assumption is reasonable \cite{ADS5400}. 
Therefore, the entire portion of the energy of $v_ {DC} (t)$ output from the rectifier can be conveyed to the energy storage and used as a DC power source without leakage to the ID. 
That is a remarkable benefit of the IntRx SWIPT architecture to enhance WPT performance.
More details of the baseband signal $y(t)$ and the demodulation process in the ID will be explained in the next section. 

\section{Modulation Design}\label{sec:modulation}
\par
This section describes the design concept and principles of modulation and demodulation of the newly proposed M-PPM for SWIPT and provides an analysis of its performance. 
The detailed method of generating the modulation signal at the transmitter and decoding the down-converted baseband signal through the EH at the IntRx SWIPT receiver's information decoder are explained using some examples and figures.
Subsequently, we analyze both power and information transfer performance using the maximum achievable data rate and the scaling laws of the energy harvester, respectively. 
%

\subsection{M-PPM Modulation and Demodulation Principle}
\par
The PPM modulation uses different positions of pulses in the time domain to modulate information as its name indicates. 
Conventional PPM modulation signal transmits a pulse for only a certain chip of time corresponding to the contained information during the entire symbol duration \cite{Carbonelli2006}, so the receiver is able to decode information by distinguishing the position of pulse and can be implemented using a simple envelope detector.
The behavioral characteristics of the envelope detector and the diode rectifier are similar, and the simple receiver design is available for PPM; therefore, those facts inspired us to apply the PPM modulation technique to IntRx SWIPT architecture in terms of WIT.
Additionally, the PPM signal waveform has a high PAPR due to its power concentration to the pulse, and the high PAPR has been widely verified as a factor that enhances the power transfer efficiency in various WPT literature. 
The M-PPM for SWIPT is proposed based on both benefits in terms of WIT and WPT to enable a low-power information decoder and, at the same time, a more efficient RF to DC power conversion at the energy harvester.
There are important considerations to enable M-PPM demodulation in the IntRx SWIPT receiver, although the modulation principle of M-PPM for SWIPT is mostly identical to the existing PPM modulation method \cite{Carbonelli2006, Garrett1983}.
The detailed modulation principle of M-PPM for SWIPT is as follows.
\par
\begin{figure}
	\centering
	\includegraphics[width=0.48\textwidth]{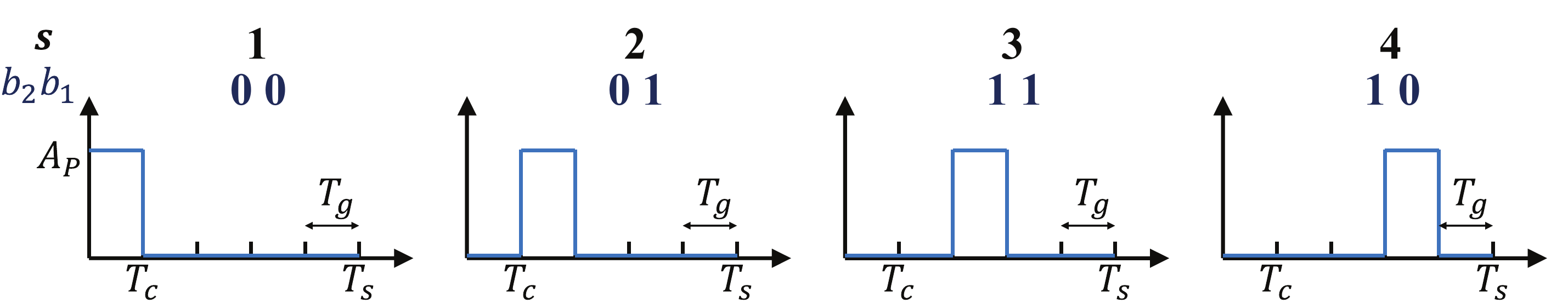}
	\caption{Examples of PPM ($M=4$) symbols and baseband signal.}
	\label{fig_symbol mapping}
\end{figure}

Each symbol of M-PPM carries a message $s$ out of $M$ possible messages, $s \in \{1,2,\ldots,M \}$, and has a duration of $T_{s}$.
The message $s$ contains $m=\log_{2}M$ bits of information that are mapped into one of $M$ symbols.
Fig. \ref{fig_symbol mapping} shows the example of modulation symbols and signals of 4-PPM, which indicates the grey coded 2-bit information is mapped into the messages $s \in \{1,\ldots , 4 \}$ and each $s$ is allocated to the respective time-delayed pulse with duration $T_{c}$. 
The $A_{p}$ stands for the amplitude of the pulses, and the average power of each symbol is equal. 
The symbol duration $T_{s}$ is divided into $M+1$ chips, with each chip representing its respective message $s$ with duration $T_{c}=T_{s}/(M+1)$, and the remaining chip is used as a guard duration $T_{g}=T_{c}$. 
This guard period $T_{g}$ is a crucial factor for M-PPM for SWIPT and contrasts with the conventional PPM, which does not require $T_{g}$. 
Unlike the conventional information receiver, the baseband signal for demodulation is obtained through a rectification process at the EH in the IntRx SWIPT receiver, and there is a smoothing circuit at the output of the rectifier to supply a stable DC power. 
At the end of the symbol period, the discharging slope of the output voltage signal at the SWIPT receiver is more flattened than the signal at the conventional information receiver, and the preceding symbol could more frequently affect the later symbol.
Therefore, $T_{g}$ is applied specifically to the modulation design for IntRx SWIPT system to reduce the inter-symbol interference (ISI) by preventing the overlap into adjacent symbols.
\par
The baseband modulated signal $x(t)$ can be modeled and represented as 
\begin{equation}\label{eq:5}
x(t) = \sum_{k=0}^{\infty}A_{p}\phi (t-kT_{s}-(s-1)T_{c}),
\end{equation}
where the index $k$ represents the $k^{th}$ symbol in the transmission sequence of symbols and index $s$ represents the message. 
$A_{p}$ is the amplitude ($A_{p}=\sqrt{2(M+1)P}$) of pulse and is limited by the average transmit power constraint $P$, and $\phi(t)$ is a rectangular pulse with unit amplitude and chip duration $T_{c}$.
Since each M-PPM symbol in the transmission sequence contains only one pulse (duration of $T_{c}$) corresponding to one of $M + 1$ slots, the PAPR of the signal is $M+1$ and the signal bandwidth is $BW = 1/T_{c} = (M+1)/T_{s}$. 
The number of possible messages $M$ and the signal bandwidth $BW$ can be selected according to the system's power transfer efficiency or spectral efficiency requirements.

\par
\begin{figure}[t]
	\centering
	\includegraphics[width=0.48\textwidth]{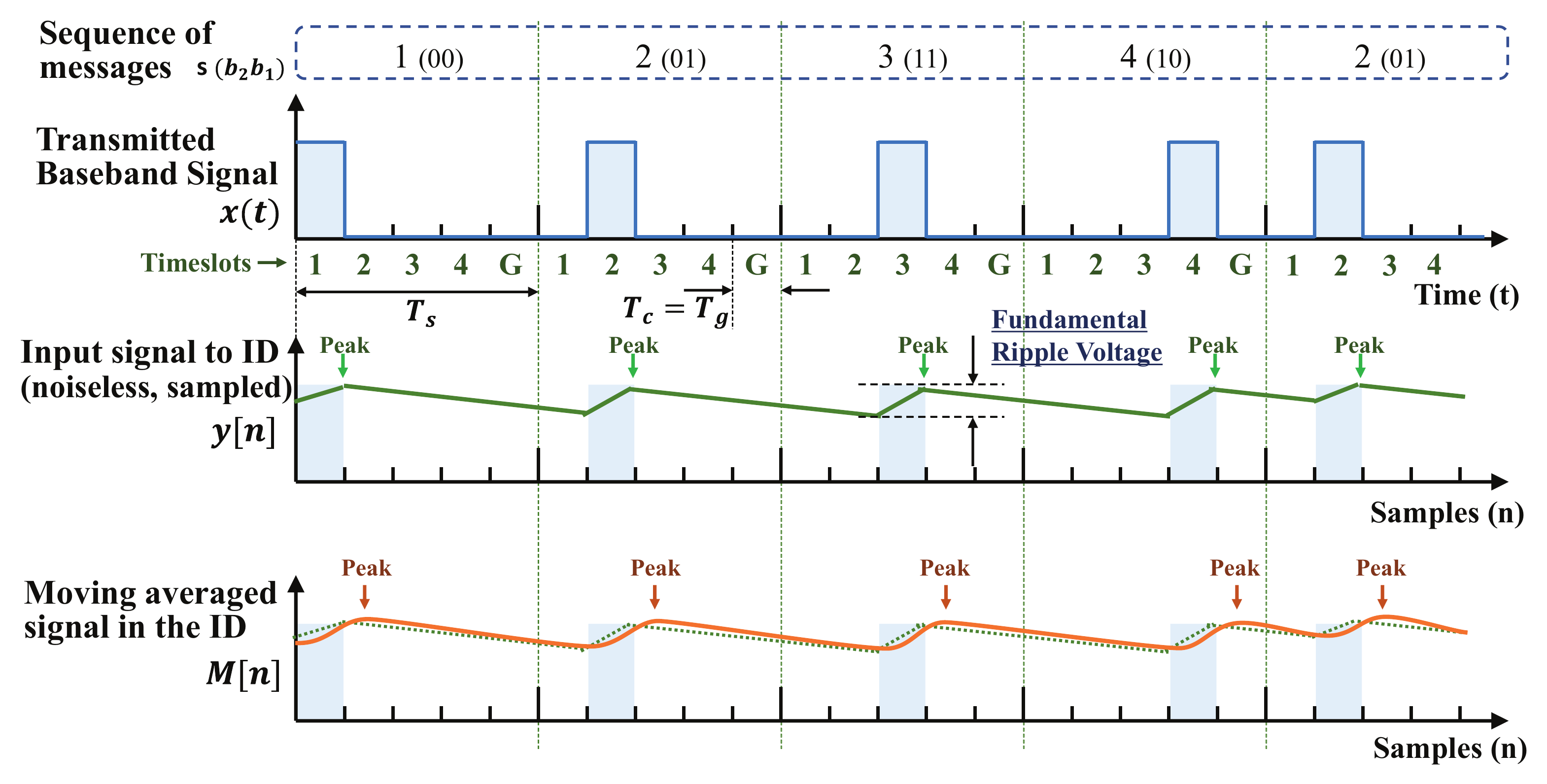}
	\caption{Example of the transmitted and received sequence of PPM symbols (M=4) and demodulation principle.}
	\label{fig_demodulation}
\end{figure}

As mentioned earlier, the input signal $y(t)$ to the ID is a low pass filtered and smoothed voltage signal obtained through EH in the IntRx SWIPT receiver, so its shape is not similar to the originally transmitted baseband signal. 
Unlike the received baseband signal of conventional PPM, distinguishing the corresponding information from the already flattened input signal is not straightforward in the IntRx SWIPT receiver. 
Nevertheless, there is a factor that enables demodulation for M-PPM for SWIPT, and we here propose a modified PPM demodulation scheme that can be applied to the IntRx SWIPT receiver architecture.
M-PPM signal has a power-concentrated pulse in a symbol duration, and the rectifier output voltage is increased during the pulse (chip) period and decreased during the rest of the time in a symbol. 
Therefore, the ripples in the output voltage signal $v_{DC}(t)$ are inevitable even without any noise. 
We call it a fundamental ripple; it is driven by the shape of the transmit waveform regardless of noise, and the size of the fundamental ripple is affected by the low pass filter constant, especially the size of the output capacitor. 
If the output capacitor size is small, the fundamental ripple increases, and the rectifier operates as an envelope detector, but if the output capacitor size becomes larger, the ripple size decreases, and the DC output signal is flattened.
The low pass filter components at the output of EH is a design factor to satisfy the system requirements for stable DC output and information transfer rate and to compensate for the performance trade-off of both aspects.
The main principle of the M-PPM demodulation scheme is that the ID is able to track the difference in voltage levels caused by the fundamental ripple over the period of one symbol and is also available to identify the contained information by seeking the peak position.
We assume perfect time synchronization; that is, the information decoder knows the starting point and the length of the information symbols from the received signals.
Fig. \ref{fig_demodulation} represents an example of the sequence of transmitted and received PPM modulated symbols in the baseband. 
Rectifier noise $n_{rect}(t)$ and RF channel effects (such as multipath fading) are not presented in the figure to explain the demodulation scheme simply and visually and focus on the role of the fundamental ripple.
\par 
A low-power digital signal processing (DSP) unit and an ADC are applied to implement M-PPM demodulation in the IntRx SWIPT receiver as shown in Fig. \ref{fig_systemmodel}.
The input signal $y(t)$ to the information decoder is digitized through the ADC and subject to the demodulation processes such as noise reduction and peak seeking in the DSP.
The $y(t)$ is sampled at the ADC by the sampling frequency of $f_{s}$, and the output of ADC can be represented as $y[n]$ where $n$ is the sampled index of the received signal $y(t)$.
As shown in the second graph of Fig. \ref{fig_demodulation}, the location of the peak voltage of ideal (noiseless) received and digitized signal $y[n]$ in a certain symbol period is always at the end of the chip that has high amplitude. This location corresponds to the information of the symbol.
Since the information decoder knows the starting point and length of the symbol and sampling frequency, the location of the peak voltage indicates what information was transmitted.
Demodulation through detecting the peak position of the ripple over a certain symbol period in a noiseless channel and rectifier circuit is straightforward.
However, if signal $y(t)$ is affected by noise or by the wireless channel, it is not possible to accurately demodulate information simply by reading and comparing the peak position of the sampled signal $y[n]$.
To resolve this issue, we have applied a digital moving average filter to reduce the effect of noise and to find the original peak position more accurately. 
The equation of moving average filter output is given by:
\begin{equation}\label{eq:5}
 M[n] = \frac{1}{L}\sum_{i=0}^{L-1}y[n-i]
\end{equation}
where $M[n]$ is an output signal of the moving average filter and $L$ denotes the number of points (samples) for averaging. 
The number of averaging samples $L$ is a design factor that can be determined by considering the signal-to-noise ratio (SNR) and the processing power of the system. 
The number of samples of the chip duration is related to the sampling rate of the ADC and can be obtained by $T_{c} \times f_{s}$. 
The third graph in Fig. \ref{fig_demodulation} shows the filtered input signal $M[n]$ which is delayed and smoother compared to the original signal $y[n]$.
We carefully set the decision boundaries to determine the message according to the peak position of the filtered signal $M[n]$ at the final stage of the demodulation process. 
The decision boundaries cannot be the same as the original time-slots in the transmission signal.
From the observation of the behavior of the received and filtered signal $M[n]$, we set the decision boundaries by delaying the chip duration of $T_{c}$, and the reason is as follows.
Firstly, the moving average filtered output $M[n]$ exhibits a delay compared to the original signal $y[n]$. 
Secondly, the peak voltage of $M[n]$ in a particular symbol is always located in the next time slot (chip) since the charging slope of the received signal is smoother than the discharging slope.
Thirdly and finally, since there is multipath fading in any real wireless channel, the delay of the moving average filtered output will be longer. 
From those reasons, we shift the decision boundaries by one next time-slot.
For example, if the peak voltage is located in time-slot two, the message is decoded as one, and if the peak voltage is located in guard time-slot, the message is decoded as $M$.
This is the reason why the guard interval $T_{g}$ is set equal to the pulse duration of $T_{c}$.
The guard duration (time-slot) is an extra time-slot that enables this shift of decision boundary, so it is essential for this modulation design. 
The information decoding procedure in the integrated receiver is summarized in the following steps:
\begin{enumerate}
\item[Step1] The received signal $y(t)$ is sampled using a sampling frequency $f_{s}$ and is converted to digital signal $y[n]$.
\item[Step2] $M[n]$ is obtained by moving average filtering the digital input signal $y[n]$ with the number of averaging sample $L$.
           $L$ is equal to the number of samples of pulse duration $L=T_{c} \times f_{s}$.
\item[Step3] Find the index of the maximum voltage from the filtered output $M[n]$.
\item[Step4] Determine the message corresponding to the index and return the bit sequence.
\end{enumerate}

\subsection{Illustration of M-PPM Demodulation}
\begin{figure}
	\centering
	\includegraphics[width=0.42\textwidth]{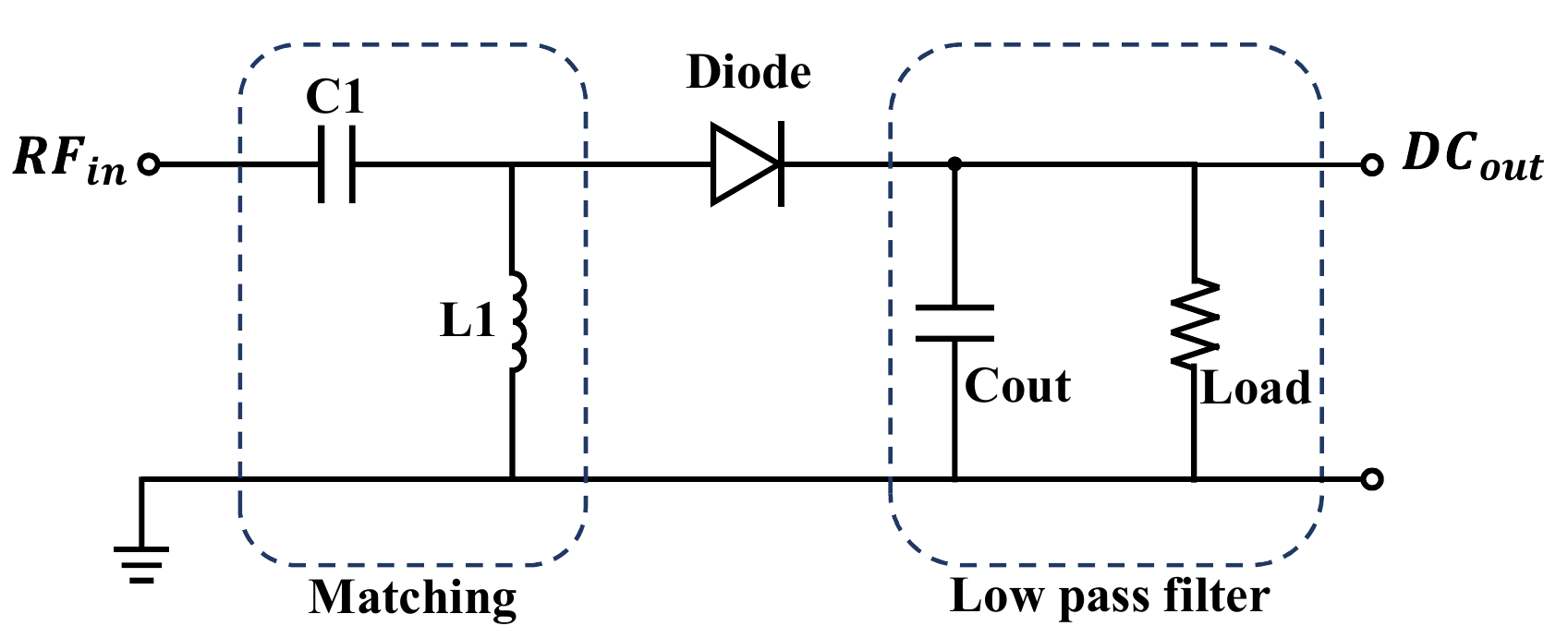}
	\caption{Circuit diagram of the single-diode rectifier (energy harvester).}
	\label{fig_circuit}
\end{figure}

We now illustrate using realistic circuit simulation the M-PPM demodulation behavior.
Since an actual energy harvester is required to describe the demodulation process and analyze the performance of M-PPM, we designed, optimized, and simulated the rectenna circuit of Fig. \ref{fig_circuit}. 
We used a conventional single series rectifier circuit that consists of a rectifying diode, impedance matching circuit, and low pass filter. 
The Schottky diode Skyworks SMS7630 is chosen for the rectifying diode because it requires a low biasing voltage level, which is suitable for a low-power rectifier. 
The impedance matching and low pass filter circuits are designed for a 2.45GHz carrier frequency with an RF input power of $-$20dBm and with wider than 10MHz bandwidth. 
The load impedance $load$ is chosen as 10k$\Omega$ in order to maximize the RF-to-DC conversion efficiency with high PAPR signals \cite{Ouda2018}. 
The matching network capacitor $C1$, inductor $L1$, and output capacitor $C_{out}$ values are optimized (using an iterative process) to maximize the output DC power under a given load impedance and for the given input waveform at $-$20dBm RF input power. 
The chosen values are given by 0.4pF for $C1$, 8.8nH for $L1$, and 1nF for $C_{out}$. 
The input signal $y(t)=v_{DC}(t)+n_{rect}(t)$ to the information decoder can be obtained at the $DC_{out}$ port in the circuit. 
The fundamental ripple levels and the slope of the charging and discharging in $v_{DC}(t)$ are varied according to the rectifier design parameters, the ripple levels in our EH circuit are only a few $mV$ when the excited RF signal power level is about $-$20 dBm at the $RF_{in}$ port. 
\par
\begin{figure}
	\centering
	\includegraphics[width=0.45\textwidth]{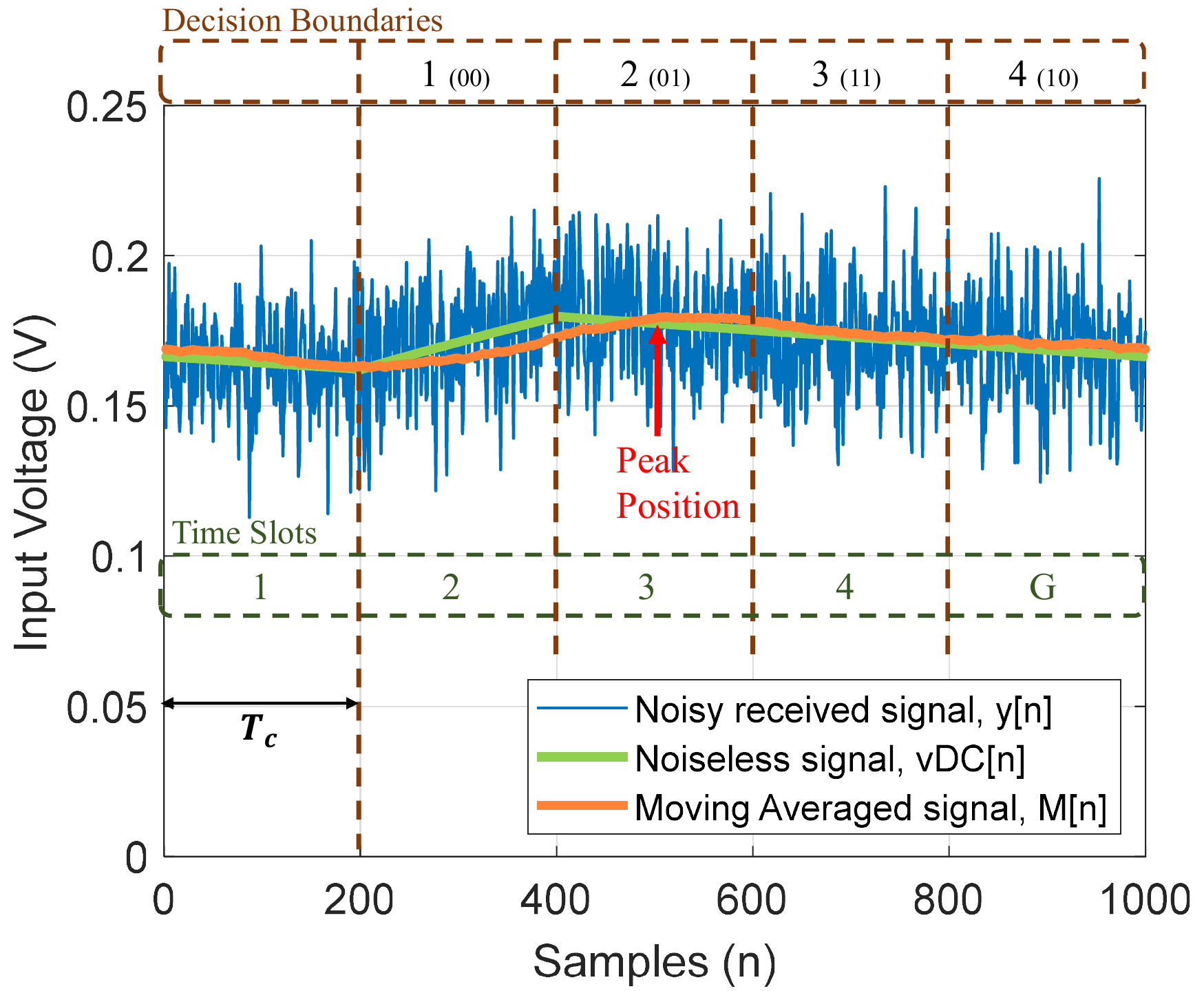}
	\caption{Example of received signal to ID with M=4, Signal Bandwidth 5 MHz, SNR 20dB, 1Gsps ADC.}
	\label{fig_rcvwave}
\end{figure}

A snapshot of an actual received 4-PPM waveform of one symbol duration is presented in Fig. \ref{fig_rcvwave}. 
The RF received signal $Y_{rf}(t)$ at the output of the AWGN channel for 4-PPM transmit signal with a bandwidth of 5MHz was generated using Matlab. The symbol duration is 1 $\mu $s, chip duration is 200 ns, and the average power of the $Y_{rf}(t)$ is $-$20 dBm.
The EH output waveform $v_{DC}(t)$ was obtained by PSpice simulations using the circuit shown in Fig. \ref{fig_circuit} and values aforementioned. 
Then, the input signal to the information decoder $y(t)$ is acquired by applying the rectifier noise $n_{rect}(t)$.
The fluctuation due to noise is shown by the blue line and is the result of applying a white Gaussian noise corresponding to 20 dB of SNR.
The input signal $y(t)$ is converted to digitised $y[n]$ with a 1 GHz sampling frequency $f_{s}$ using Matlab.
1000 samples in the x-axis represent a symbol duration of 1 $\mu$s, and each chip duration is represented by 200 samples.
The blue line in Fig. \ref{fig_rcvwave} represents the digitised input signal $y[n]$.
For reference, we also indicate the digitized noiseless EH output signal $v_{DC}[n]$ as a green line.

The moving average filtered signal $M[n]$ is shown as the orange line in the figure, where the moving average window size of the filter $L$ was set to 200 which is the same length of the chip duration.
%
%
The original message $s$ of the transmit signal was $s=2$ and is represented by bits 01. 
So, the voltage of the noiseless signal $v_{DC}[n]$ increases only during time-slot 2, and the peak position of the signal is the same as the endpoint of time-slot 2.
The M[n] signal, after noise reduction, has a similar pattern to the noiseless signal $v_{DC}[n]$ with a slight delay.
The peak voltage value of $ M [n] $ is at the 512th sample, which is located in time-slot 3.
As described in the previous section, the decision boundaries were set by shifting to the next time-slot.
In this example, the decision boundaries were set as an illustrated in Fig. \ref{fig_rcvwave}. 
Finally, the message is decoded as $s=2$ (bits of 01) which is the same as the originally transmitted message $s$.
%
%
More discussions on the information transfer performance are provided in the next subsection.

\subsection{Theoretical Performance Analysis of M-PPM in SWIPT}
\subsubsection{Information Transfer Performance} $ $
\par
\begin{figure}[t]
	\centering
	\includegraphics[width=0.4\textwidth]{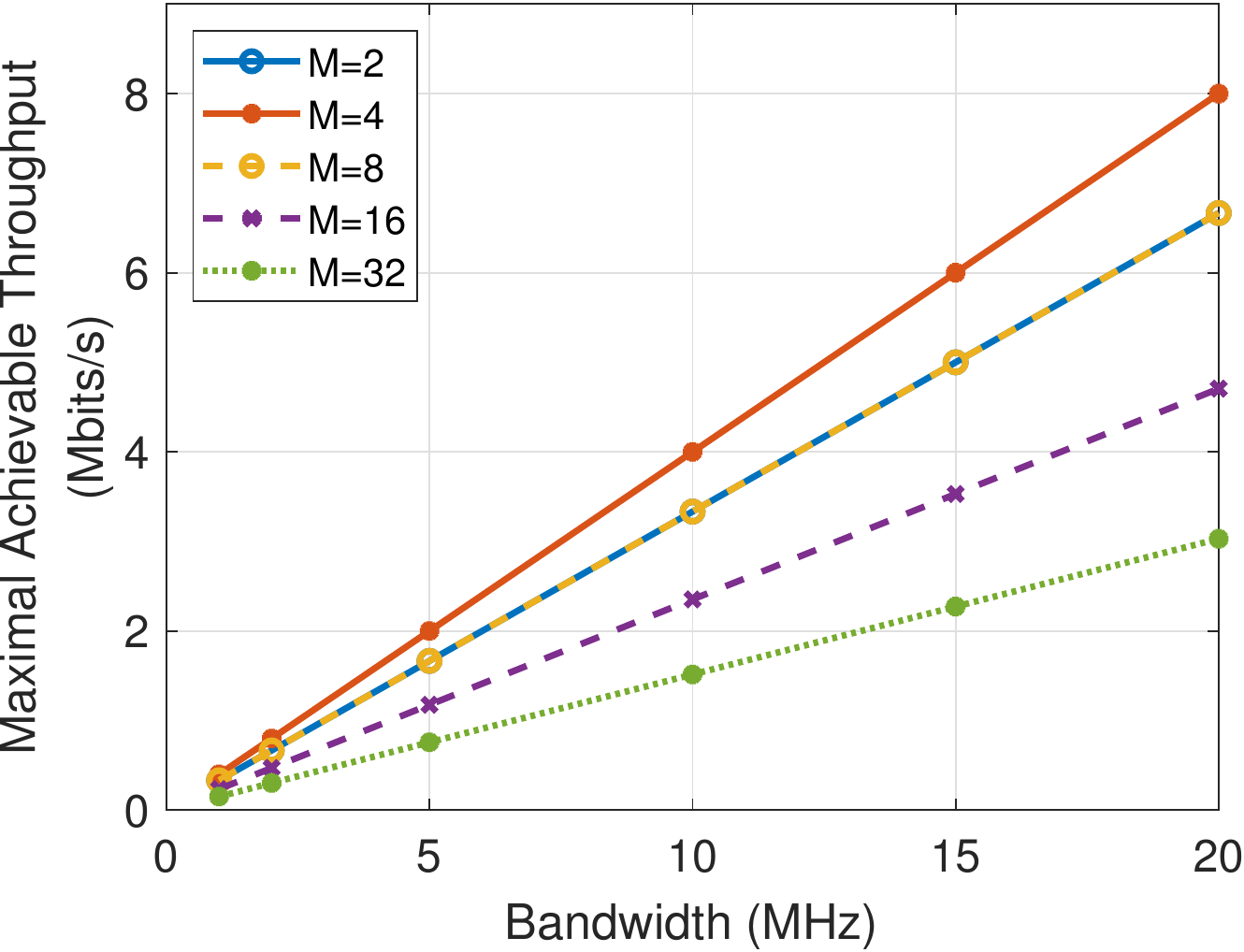}
	\caption{Throughput of different $M$ versus bandwidth}
	\label{fig_throughput}
\end{figure}

 
The data throughput $R$ of the M-PPM modulation can be determined by the signal bandwidth ($BW$) and the number of messages $M$,
\begin{equation}\label{eq:6}
 R = \frac{BW}{(M+1)}\log_2 M.
\end{equation}
The throughput of several different $M$ versus bandwidth is displayed in Fig. \ref{fig_throughput}.
It appears that a suitable choice of the number of messages $M$ is four as it leads to larger throughput compared to other $M$ displayed in Fig. \ref{fig_throughput}.
Similar to conventional communication signals, as the bandwidth increases, the throughput also proportionally increases, and the spectral efficiency of 4-PPM, a suitable choice to maximize throughput, is 0.4 bits/s/Hz. 
The spectral efficiency of the M-PPM modulation is smaller than that offered by the conventional PSK modulation schemes (BPSK has 1 bits/s/Hz of spectral efficiency).
However, the data rate generally required for low-power SWIPT receivers is not large, and RF elements such as local oscillators and mixers that are responsible for a large part of the power consumption in the conventional RF receiver are not required for information decoding in the proposed architecture. 
Consequently, M-PPM modulation has a significant practical advantage for practical SWIPT systems.

\par
\begin{figure}
	\centering
	\includegraphics[width=0.48\textwidth]{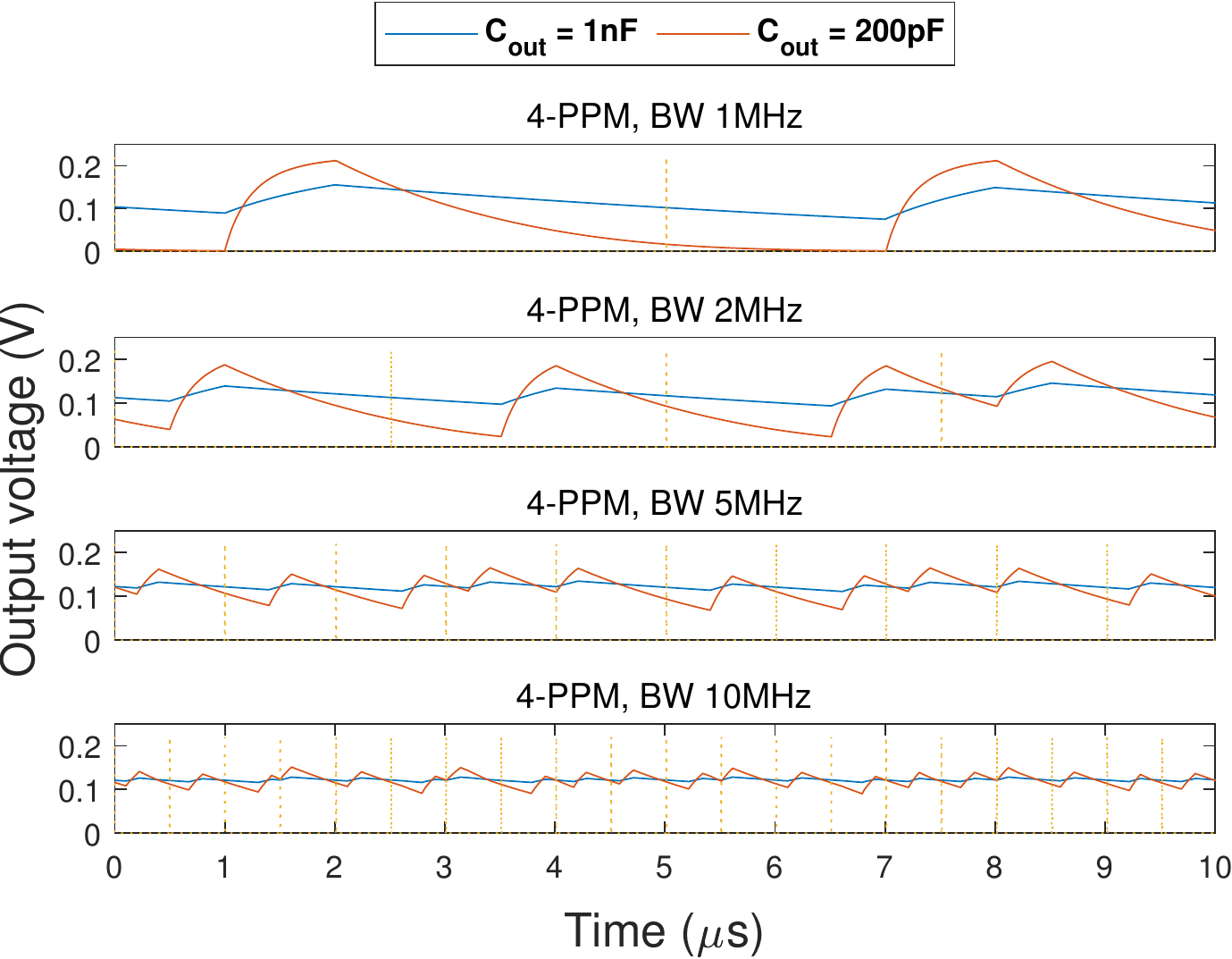}
	\caption{Fundamental ripples occured by 4-PPM input signals with different bandwidth and $C_{out}$ of EH (input power of -20dBm).}
	\label{fig_fund_ripple}
\end{figure}

The M-PPM modulation's actual throughput performance for SWIPT is highly related to the system characteristics and hardware configurations.
The received M-PPM modulated signal at the EH induces a slope of the charging and discharging in $v_{DC}(t)$ output signal during the rectification process. 
We earlier defined the fundamental ripple of the received M-PPM signal, which is a slope characteristic that is dictated only by the input waveform rather than by the noise at the rectifier.
Recall that the ripple voltage at the output of the EH is a key element of this modulation and demodulation technique.
The fundamental ripple is largely dependent on the hardware components that make up of the EH (see Fig. \ref{fig_circuit}) such as load, and output capacitor ($C_{out}$), and is also dependent on the signal characteristics such as input power, bandwidth, and $M$.
For instance, for the design characteristics in the previous subsection such as $load$ = 10k$\Omega$ and $C_{out}$ = 1nF, the peak-to-peak fundamental ripple voltage is only a few $mV$ when the signal characteristics are $BW=1MHz$, $M=4$ and the excited RF signal power level is about -20 dBm at the receiving antenna.
In the actual received signal $y(t)$, a ripple caused by rectenna noise and the wireless channel fading is added to the fundamental ripple.
If the noise ripple voltage is larger than the peak-to-peak of the fundamental ripple voltage, it is difficult to seek the peak position of the fundamental ripple accurately at the ID.
So, the SNR of $y(t)$ is also closely linked to the proportion of the noise ripple and the fundamental ripple in $v_{DC}(t)$.
When the $C_{out}$ value and bandwidth are set large and $M$ is set small, $v_{DC}(t)$ becomes smoother, the peak-to-peak fundamental ripple voltage decreases, which causes a decrease of the SNR of $y(t)$ even if the noise power at the rectifier is not changed.
On the other hand, if $C_{out}$ and bandwidth are set too small and $M$ becomes large, the fundamental ripple voltage is increased which improves the SNR, but the output waveform shape gets closer to a sawtooth which makes it difficult to supply stable DC power to the power management module. 
Fig. \ref{fig_fund_ripple} displays the fundamental ripples of $v_{DC}(t)$ obtained from the 4-PPM signals with several different bandwidths ($BW$ = 1, 2, 5, and 10 MHz) and two different $C_{out}$ values (200pF, 1nF) of EH.
These values of bandwidth and $C_{out}$ will be used for the numerical and experimental performance evaluation of information transfer in the rest of this paper. 
These trade-offs between SNR and DC stability due to rectifier design and signal characteristics should be carefully considered to meet the requirements of the SWIPT system.
Not only the EH hardware and signal design but also ID configuration have a significant effect on information transfer performance.
The sampling frequency $f_{s}$ and resolution of the ADC determines the number and the precision of digitized samples of the symbol duration.
That is, the higher the sampling frequency $f_{s}$, the more data samples can be acquired from the input signal $y(t)$.
It allows the demodulator to have more room for adjusting the moving average window size $L$ and is helpful to reduce the effect of noise during demodulation.
In addition, if a high-resolution ADC is used, the input signal's voltage values can be more precisely obtained even if the peak-to-peak voltage of the ripple is very small and the performance of the information demodulation can be enhanced.
Due to those many different design factors such as $C_{out}$, $BW$, $M$, $f_{s}$, and $L$, an analysis of the information transfer performance of the M-PPM modulation technique is challenging. 
So, we provide an evaluation of the information transfer performance numerically and experimentally in the rest of this paper. 

\subsubsection{Power Transfer Performance} $ $
\par
In \cite{Clerckx2016}, the rectifier output current (and therefore power) was modeled using a nonlinear diode model based on the Taylor expansion of a diode characteristic function truncated to the fourth moment, and the power transfer performance was analyzed.
We here reuse the same metric and model the average delivered power (denoted by $P_{del}$) as
\begin{equation}\label{eq:4}
 P_{del} = \mathbb{E}\{k_{2}R_\mathrm{ant}Y_{\mathrm{RF}}(t)^{2}+k_{4}R_\mathrm{ant}^{2}Y_{\mathrm{RF}}(t)^{4} \},
\end{equation}
where $k_{2}$ and $k_{4}$ are constants values of 0.0034 and 0.3829 respectively, $R_\mathrm{ant}$ is an antenna impedance and $\mathbb{E}\{ .\}$ is an expectation and averaging operator over the randomness of the signal and over time.
Since there is no leakage to the ID in the IntRx SWIPT receiver, $P_{del}$ is the same as the average received DC power of the EH receiver. 
In this section, we derive the theoretical scaling laws of $P_{del}$ for the novel M-PPM modulated signal.
The transmission is assumed narrowband, and the channel is assumed frequency flat. 
Following \cite{Clerckx2016}, we can write
\begin{align}
 P_{del} &= k_{2}R_\mathrm{ant}\mathbb{E}\{Y_{\mathrm{RF}}(t)^{2}\} + k_{4}R_\mathrm{ant}^{2}\mathbb{E}\{Y_{\mathrm{RF}}(t)^{4}\} \nonumber \\
 &= k_{2}R_\mathrm{ant}\mathbb{E}\{\left|m(t)\right|^{2}\}P + \frac{3}{2}k_{4}R_\mathrm{ant}^{2}\mathbb{E}\{\left|m(t)\right|^{4}\}P^{2}\nonumber \\
 &= k_{2}R_\mathrm{ant}P + \frac{3}{2}k_{4}R_\mathrm{ant}^{2}\mathbb{E}\{\left|m(t)\right|^{4}\}P^{2}
\end{align}
where $m(t)$ is the normalised modulations symbol. The $M$ possible symbol waveforms are : $m(t) = A_{p}\phi (t-sT_{c})$ with $s \in \{0, 1, \ldots,M-1 \}$. 
Note that the signal structure of M-PPM signal is similar\footnote{Each M-PPM signal can also be expressed according to the OOK signaling of \cite{Varasteh2020} with $l=\sqrt{M+1}$, for example 2-PPM, 4-PPM, and 8-PPM signal's $l$ parameters are $\sqrt{3}$, $\sqrt{5}$, and $\sqrt{9}$ respectively.} to the OOK signaling in \cite{Varasteh2020}, but the probability of high amplitude pulse is fixed according to the number of symbols $M$. 
Table \ref{table_scalinglaw} displays the $P_{del}$ of several M-PPM modulations and conventional waveforms (continuous wave (CW), PSK, and QAM) as baseline.
%
\begin{table}
\centering
\caption{Scaling Laws of M-PPM and Conventional Modulations}
\label{table_scalinglaw}
\renewcommand{\arraystretch}{1.2}
{\small
\begin{tabular}{>{\centering}m{25mm} >{\centering}m{40mm}}
 \toprule
 Modulation			& $P_{del}$ 									 \tabularnewline \midrule[1.5pt] 
Continuous Wave (CW)	& $k_{2}R_\mathrm{ant}P + 1.5k_{4}R_\mathrm{ant}^{2}P^{2}$     \tabularnewline \midrule
BPSK				& $k_{2}R_\mathrm{ant}P + 1.5k_{4}R_\mathrm{ant}^{2}P^{2}$     \tabularnewline \midrule
16QAM				& $k_{2}R_\mathrm{ant}P + 1.98k_{4}R_\mathrm{ant}^{2}P^{2}$     \tabularnewline \midrule
2-PPM 				& $k_{2}R_\mathrm{ant}P + 4.5k_{4}R_\mathrm{ant}^{2}P^{2}$			\tabularnewline	 \midrule
4-PPM 		     		& $k_{2}R_\mathrm{ant}P + 7.5k_{4}R_\mathrm{ant}^{2}P^{2}$		\tabularnewline	 \midrule
8-PPM 				& $k_{2}R_\mathrm{ant}P + 13.5k_{4}R_\mathrm{ant}^{2}P^{2}$		\tabularnewline	 \midrule
\bottomrule
\end{tabular}
}
\end{table}
%
\par
Table \ref{table_scalinglaw} shows that the coefficient of the second-order term is the same for all different signals. 
However, the M-PPM signals exhibit larger coefficients of the fourth-order term as expected due to its waveform shape that has a large amplitude pulse for a short period and no signals during the rest of the period in the symbol duration. 
In other words, the M-PPM boosts the fourth-order moment of the input signal $\mathbb{E}\{\left|m(t)\right|^{4}\}$ which is known to improve the RF-to-DC conversion efficiency and the power transfer performance \cite{Clerckx2016, Clerckx2019, Clerckx2018, Clerckx2018a, Varasteh2020}. 
This shows that the power transfer performance can be enhanced using M-PPM over conventional modulations.
%

\section{Numerical Performance Evaluation}\label{sec:simulation} 
\par
We evaluate the power transfer and the information transfer performance of M-PPM using circuit simulation. 
The wireless channel is assumed to be AWGN and the rectifier noise $n_{rect}(t)$ is Gaussian distributed around the rectifier output voltage $v_{DC}(t)$.
We consider M-PPM with a signal bandwidth of 1, 2, 5, and 10 MHz correspondings to a pulse duration of 1$\mu$s, 500ns, 200ns, and 100ns, respectively. 
The rectifier circuit and component values for the simulation are the same as the ones shown in Fig. \ref{fig_circuit} and the previous section (the values are 0.4pF for $C1$, 8.8nH for $L1$, 10k$\Omega$ for $load$, and 1nF for $C_{out}$).
Additionally, we also provide simulation results for $C_{out}$ equal to 200pF to illustrate the impact on information and power transfer performance. 
The input RF power to the receiver is always fixed to -20dBm level, and the sampling rate of the ADC at the ID is set to 2 Giga samples per second (GSPS).
Any channel coding or pulse shaping is not applied to evaluate the performance of the modulation scheme itself. 

\subsection{Information Transfer Performance}
\begin{figure}
	\centering
	\includegraphics[width=0.4\textwidth]{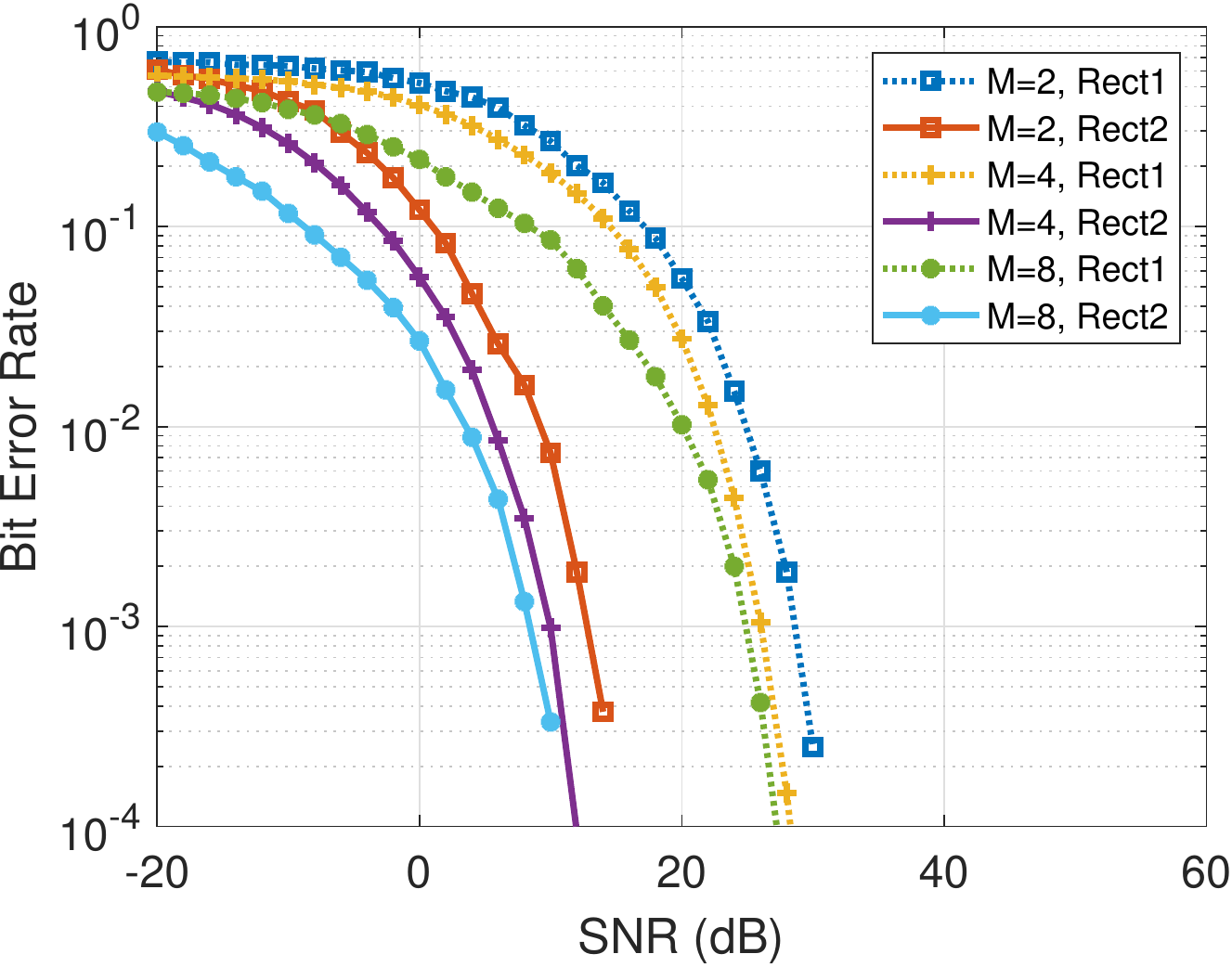}
	\caption{Simulated BER performance versus SNR of M-PPM signals with 5MHz bandwidth and three different modulation order (M = 2, 4, 8), Rect1 : $C_{out}$=1nF, Rect2 : $C_{out}$=200pF.}
	\label{fig_ppm_sim_ber_M}
\end{figure}

To analyze the information transfer performance of M-PPM modulation, we simulated the BER performance of the M-PPM signals with different design parameters.
The BER simulation was carried out using jointly Matlab and Pspice software.
M-PPM modulated RF signals are generated in Matlab, and the signals are passed to the Pspice for the rectifier circuit simulation, then the circuit simulation output voltage data $v_{DC}(t)$ way back to Matlab to apply noises and to be demodulated, and finally BER is calculated by comparing the demodulated signal and the original signal in Matlab. 
Fig. \ref{fig_ppm_sim_ber_M} shows the BER performance simulation results versus SNR with various modulation order ($M$), and different rectifier design ($C_{out}$), and fixed bandwidth of 5 MHz. 
\par
The graph in Fig. \ref{fig_ppm_sim_ber_M} shows that the BER performance tends to slightly improve as the modulation order $M$ increases for both rectifier designs.
When the transmit power and bandwidth are fixed and we increase the modulation order, the symbol duration is increased but the length of the pulse (chip) does not change due to the fixed bandwidth.
According to the principle of M-PPM modulation, the symbol duration is ($M+1$) times the pulse length, and the amplitude of the pulse also increases with increasing $M$ to maintain the average power of the M-PPM symbol (refers to \eqref{eq:5}). 
Accordingly, a higher-order PPM signal transmits a pulse of larger amplitude and does not transmit any signals for the longer periods of time of $\frac{M}{M+1} \times T_{c}$, and results in a larger fundamental ripple at the receiver.
A larger fundamental ripple is one of the causes of improving BER performance with higher-order modulation because the larger peak-to-peak voltage of the fundamental ripple offers more immunity to the same level of noise.
The other reason for improving BER performance with higher-order modulation is the grey-coded bit mapping. 
Many errors occurred in the adjacent time slot, and the difference between the adjacent slot is only one bit due to the grey code mapping, so, higher-order code can decrease error rate by $\frac{1}{\log_{2}M}$ when the error occurred in the adjacent slot. 
Although a higher-order PPM modulation improves the BER performance slightly, the highest throughput can be achieved with 4-PPM as we analyzed in the previous section.
%
%
From the simulation results, BER of $10^{-4}$ can be achieved at approximately 30 dB SNR, and the data throughput is 1.6665, 1.9998, and 1.6665 Mbps for M = 2, 4, and 8, respectively, when using an EH with 1nF output capacitor and the signal bandwidth of 5 MHz.
The same performance can be achieved at around 10 dB SNR by using an EH with a 200pF output capacitor.
These results also clearly show that a better BER performance can be obtained using the rectifier with a smaller $C_{out}$ value. 
A larger $C_{out}$ value of the rectifier causes a longer charging and discharging time of the capacitor and makes the output signal smoother. 
A smoother voltage output corresponds to a lower fundamental ripple (see Fig. \ref{fig_fund_ripple}), which leads to worse BER performance. 

\par
\begin{figure}[t]
	\centering
	\includegraphics[width=0.4\textwidth]{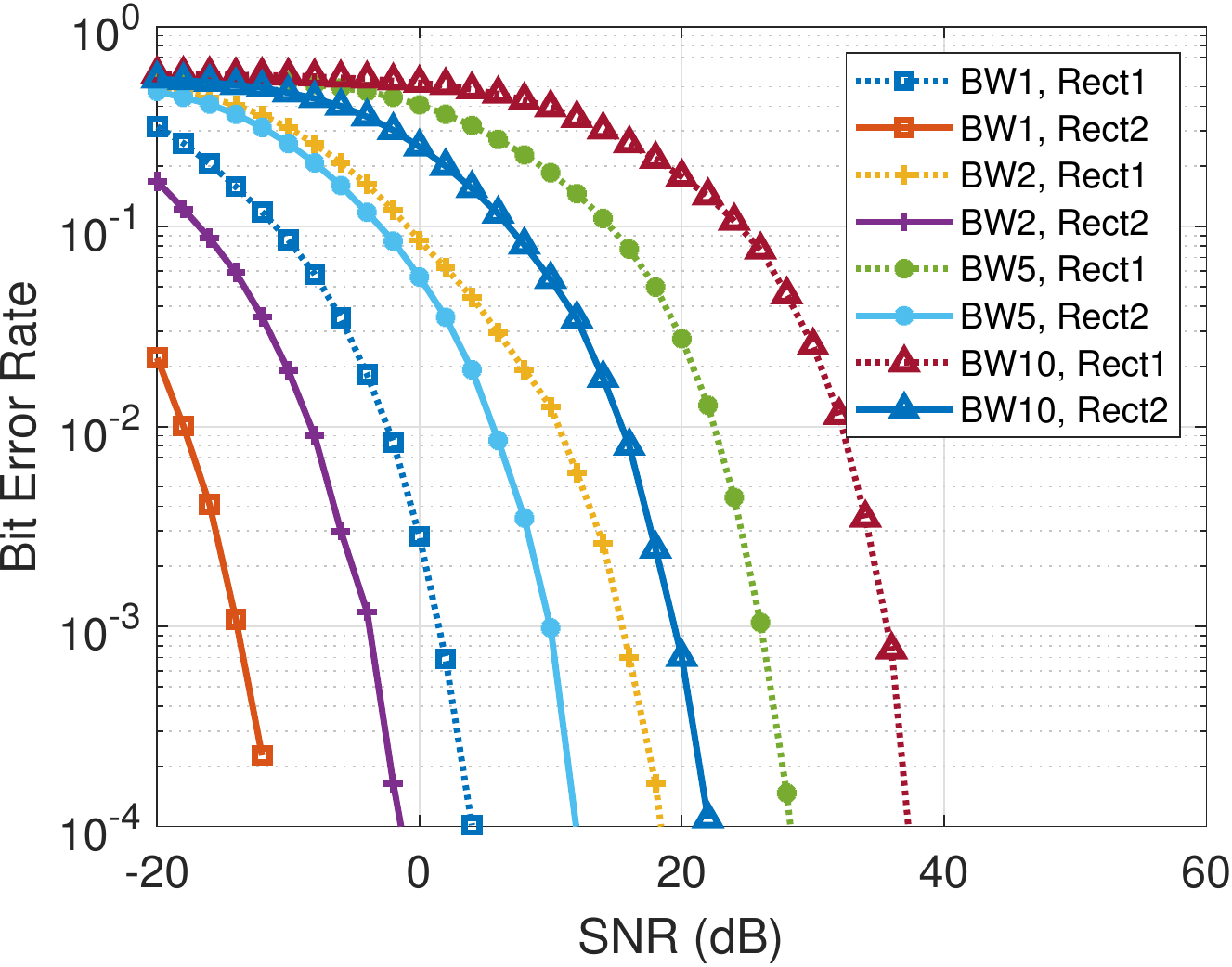}
	\caption{Simulated BER performance of M-PPM signals with different bandwidth (BW at MHz), Rect1 : $C_{out}$=1nF, Rect2 : $C_{out}$=200pF.}
	\label{fig_ppm_sim_ber_BW}
\end{figure}

Fig. \ref{fig_ppm_sim_ber_BW} displays the BER performance of M-PPM signals with various bandwidth (BW) and fixed modulation order of $M$=4. 
The graph shows that larger bandwidth signals tend to be more vulnerable to noise.
Those results are also related to the fundamental ripple of each PPM signal with different bandwidth. 
A higher bandwidth of the M-PPM signal has a shorter length of the pulse (chip) duration and symbol duration. 
When the rectifier design parameters are fixed, a shorter symbol duration makes a more flattened output DC voltage signal, and the fundamental ripple decreases (recall the Fig. \ref{fig_fund_ripple}).
For example, the fundamental ripple for a signal with M = 4 with 10 MHz BW is 13 mV and the signals with M = 4 with 5 MHz BW is 22 mV when using $C_{out}$ value of 1nF at the rectifier.
The interval between the pulses of the 5 MHz signal and the 10 MHz signal is different, and the 5MHz signal has a longer interval, therefore, the discharging time of $C_{out}$ is larger. 
Moreover the charging time of $C_{out}$ is the same as the pulse duration $T_{c}$, and the 5MHz signal has a twice longer charging time. 
The BER performance is, of course, better with the 5MHz BW signal because the fundamental ripple is about twise larger than that of a 10MHz BW signal.
Due to the longer charging and discharging time, the smaller BW signal shows a higher fundamental ripple and stronger noise immunity than larger BW signals, and it is beneficial for BER performance. 
However, the data throughput of the M-PPM modulation is proportional to the signal bandwidth, so the smaller bandwidth for improving BER performance causes a data throughput decrease.
For instance, when the SNR is 30dB, the BER of 4-PPM with 5 MHz bandwidth signal is less than $10^{-5}$ and 10 MHz bandwidth signal shows $10^{-2}$ of error rate. 
Then, the data throughput of each signal can be calculated as 1.99 Mbps and 3.89 Mbps for the bandwidth of 5 MHz and 10 MHz, respectively.
The fundamental ripple size can also be adjusted by modifying a rectifier hardware design. 
As illustrated in the graph, the rectifier with smaller output capacitor $C_{out}$ outperforms the bigger one in terms of BER performance in all the test cases with the same bandwidth BW PPM signal. 
In some cases, even the simulation results with a smaller bandwidth signal and larger capacitor rectifier are not as good as the results with larger bandwidth and a smaller capacitor. 
\par 
Those performance behaviors with different modulation index $M$ and bandwidth show that the signal design parameters cause a trade-off between data throughput and BER performance, somewhat similar to conventional modulation schemes. 
The benefit of the M-PPM modulation for SWIPT is that the performance can be tuned by simply changing system design parameters such as $C_{out}$ of the EH. 
Exploiting these trade-offs and relations by adjusting the bandwidth, modulation order, and energy harvester design, a SWIPT system with M-PPM modulation can be optimized to meet various requirements.
\par 
Apart from the aforementioned signals and system design parameters, the resolution, sampling rate of the ADC, and the moving average filter's sample size could also affect the BER performance of this modulation technique.
Indeed, we can simply expect that more data points acquired through an ADC with high resolution and sample rate and a larger size of the filter window $L$ will further reduce the impact of noise and result in a better BER performance.
Also, there could be some trade-offs for system performances, such as each component's power consumption. 
However, in this paper, we focus only on the three factors mentioned earlier (modulation order, bandwidth, and EH design) and do not consider many other factors related to ADC and filter.
Those are left for future work. 

\par
This section only discussed information transfer performance, so only the impact of the factors such as bandwidth, modulation index, and EH design on the BER and throughput performance was investigated. 
From these discussions, reducing the capacitor size $C_{out}$ on the EH can be acknowledged as a parameter that improves the SWIPT system performance.
The simulation results provided in this section definitely show better system performance using smaller $C_{out}$ in the information decoder perspective.
However, it does not guarantee a better system design in terms of the energy harvester performance.
In the next section, we evaluate the power transfer performance of the M-PPM modulation signal and discuss the trade-off between information and power transfer.
%
\subsection{Power Transfer Performance}
\begin{figure}[t]
	\centering
	\subfigure[]{\includegraphics[width=0.22\textwidth]{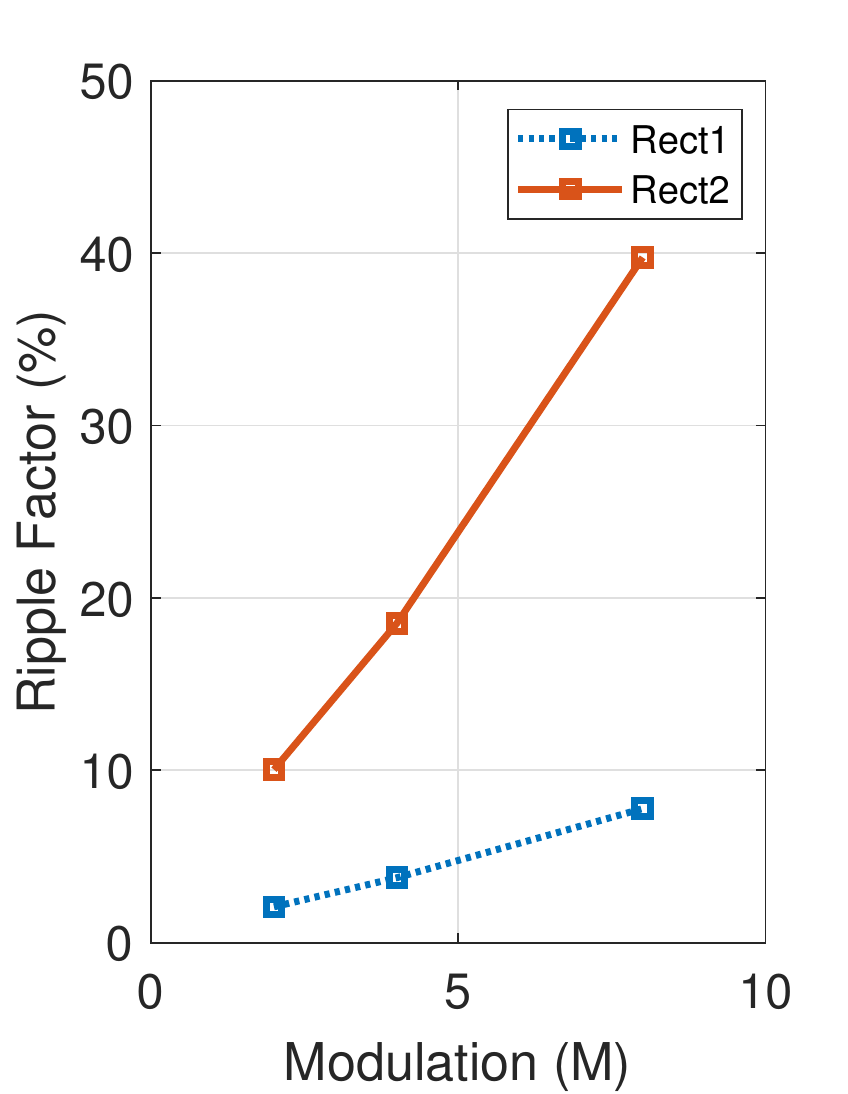}}
	\subfigure[]{\includegraphics[width=0.22\textwidth]{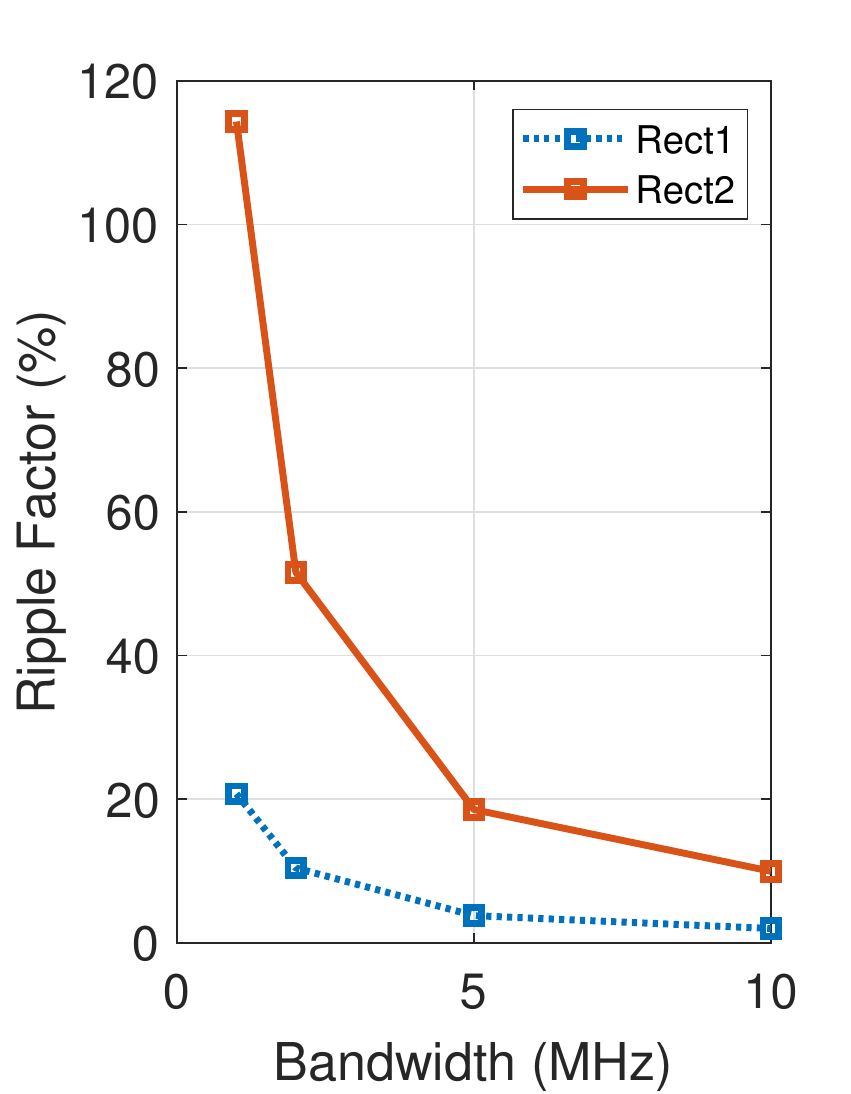}}
	\caption{Fundamental ripple factors of M-PPM modulated signals (a) versus modulation order (b) versus bandwidth, Rect1 : $C_{out}$=1nF, Rect2 : $C_{out}$=200pF.}
	\label{fig_ripplefactor}
\end{figure}

In the previous section, we discussed many signals and system design parameters that affect information transfer significantly. 
To improve the BER performance, it is confirmed that using a higher-order modulation, a smaller bandwidth, and a smaller capacitor on the rectifier is a good solution. 
All those parameters increase the fundamental ripple of the PPM modulated signal, and it is beneficial to clearly identify the peak position of the signals, which relates to the contained information. 
However, in the SWIPT system, power transfer is also an important performance factor. 
If the signal and system parameters are designed only to improve the information transfer performance by increasing the ripple of the EH output signal, a stable DC power cannot be supplied.
The output signal can be a pulse chain or sawtooth-shaped signal that is not efficient to use or store as DC power. 
The ripple occurred by noise is inevitable, but the fundamental ripple can be adjusted by modifying signal and system parameters. 
To evaluate the quality of the DC output of the EH based on a fundamental ripple, we use a ripple factor, which is the ratio of the AC and DC components in the voltage signal.
Fig. \ref{fig_ripplefactor} shows fundamental ripple factors of several noiseless M-PPM signals at the EH output with different bandwidth, modulation order, and rectifier design. 
As confirmed earlier, Fig. \ref{fig_ripplefactor} indicates that the ripple factor increases significantly in systems with a smaller bandwidth, a higher modulation order, and small rectifier output capacitors, which are factors that improve information transfer performance.
Conversely, larger bandwidth and rectifier output capacitor lead to stable DC output with a smaller ripple factor. 
\par
\begin{figure}[t]
	\centering
	\includegraphics[width=0.4\textwidth]{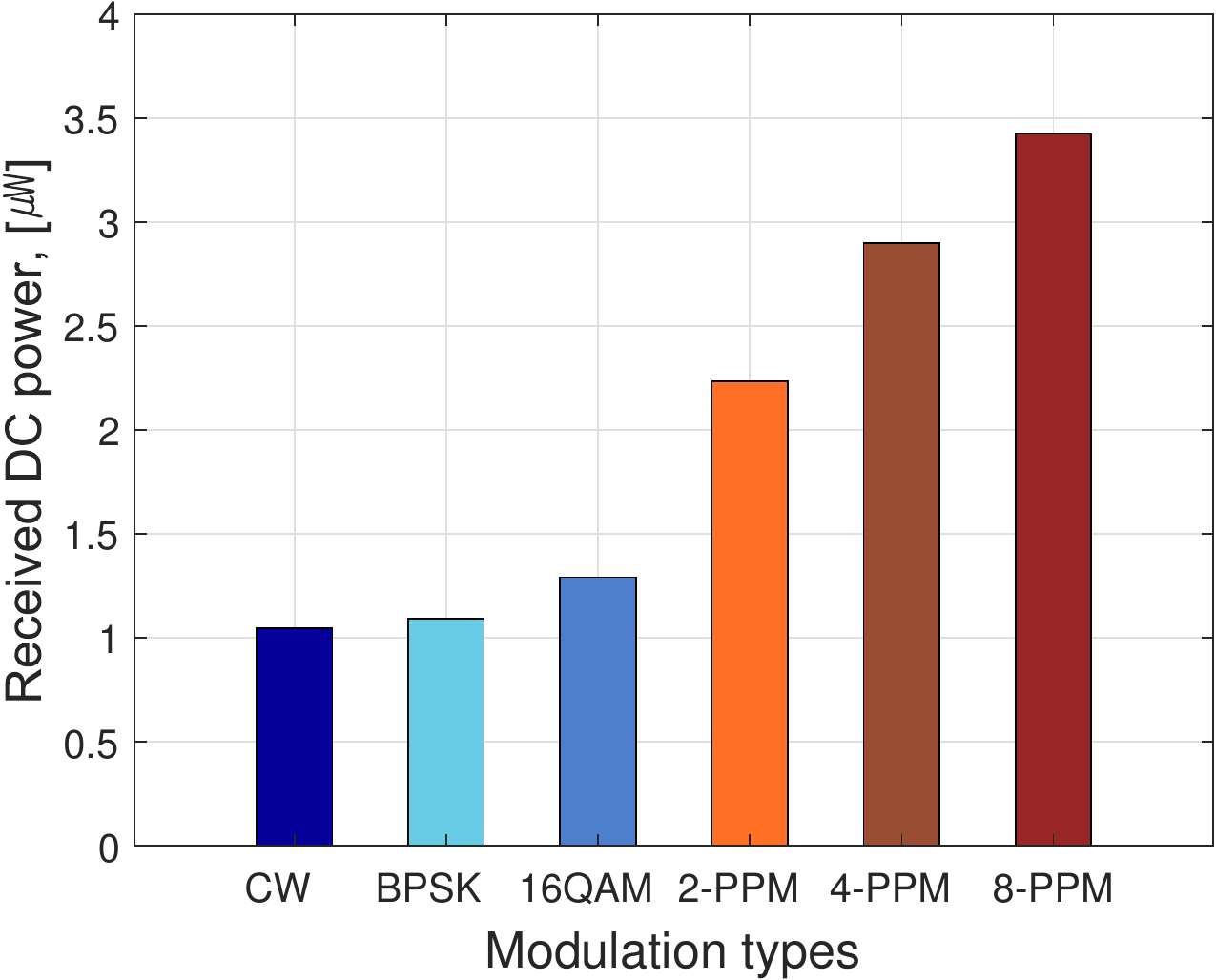}
	\caption{Harvested DC power performance with M-PPM signals with -20dBm input RF power.}
	\label{fig_ppm_sim_power}
\end{figure}

We select the MPPM signal with 5 MHz bandwidth and the 1nF output capacitor of the rectifier ($C_ {out} $) to evaluate the power transfer performance based on its relatively low ripple factor.
In Table \ref{table_scalinglaw}, the fourth-order term of the M-PPM signals is shown to be large, so the power transfer performance of the M-PPM signals is expected to outperform the normal CW and conventional modulation signals. 
As we already have seen from the analytical, simulation and measurement results of various modulation signals in \cite{Kim2020}, the relationships between theoretical $z_{\mathrm{DC}}$ and actual performance are well confirmed. 
The power transfer performance simulation results and their comparison to existing modulation schemes are shown in Fig. \ref{fig_ppm_sim_power}. 
The average harvested power simulation was carried out using 2, 4, 8-PPM with 5 MHz bandwidth signal with its input power of -20dBm, and using the rectifier with the $C_{out}$ value of 1nF.
\par
As we expected in Table \ref{table_scalinglaw}, the simulation results show that M-PPM signals outperform the conventional modulation signals such as PSK and QAM as well as CW. 
The results clearly indicate the significant gains of M-PPM signals over CW and conventional modulations. 
2-PPM shows 114\% of the gain, and 8-PPM shows more than 200\% of the gain over CW in the simulation. 
Those big gains could be achieved thanks to the large fourth-order term offered by PPM signals in $z_{\mathrm{DC}}$ characteristics, as well as our rectifier circuit design characteristics. 
The large fourth-order term of $z_{\mathrm{DC}}$ accounts for the nonlinear characteristics of the diode.
The transmission pulses of M-PPM modulation have a higher amplitude and a lower occurrence probability when M is increasing, which increases the PAPR, the fourth-order term in $z_{\mathrm{DC}}$, and consequently the power transfer performance.
%
%
\section{Experiments and Validation}\label{sec:experiment}
In this section, we establish a testbed for the SISO WIPT system in an indoor office environment and evaluate information and power transfer performance using PPM modulation signals. 

\subsection{WIPT System Implementation and Testbed Setup}

\begin{figure}[t]
	\centering
	\includegraphics[width=0.48\textwidth]{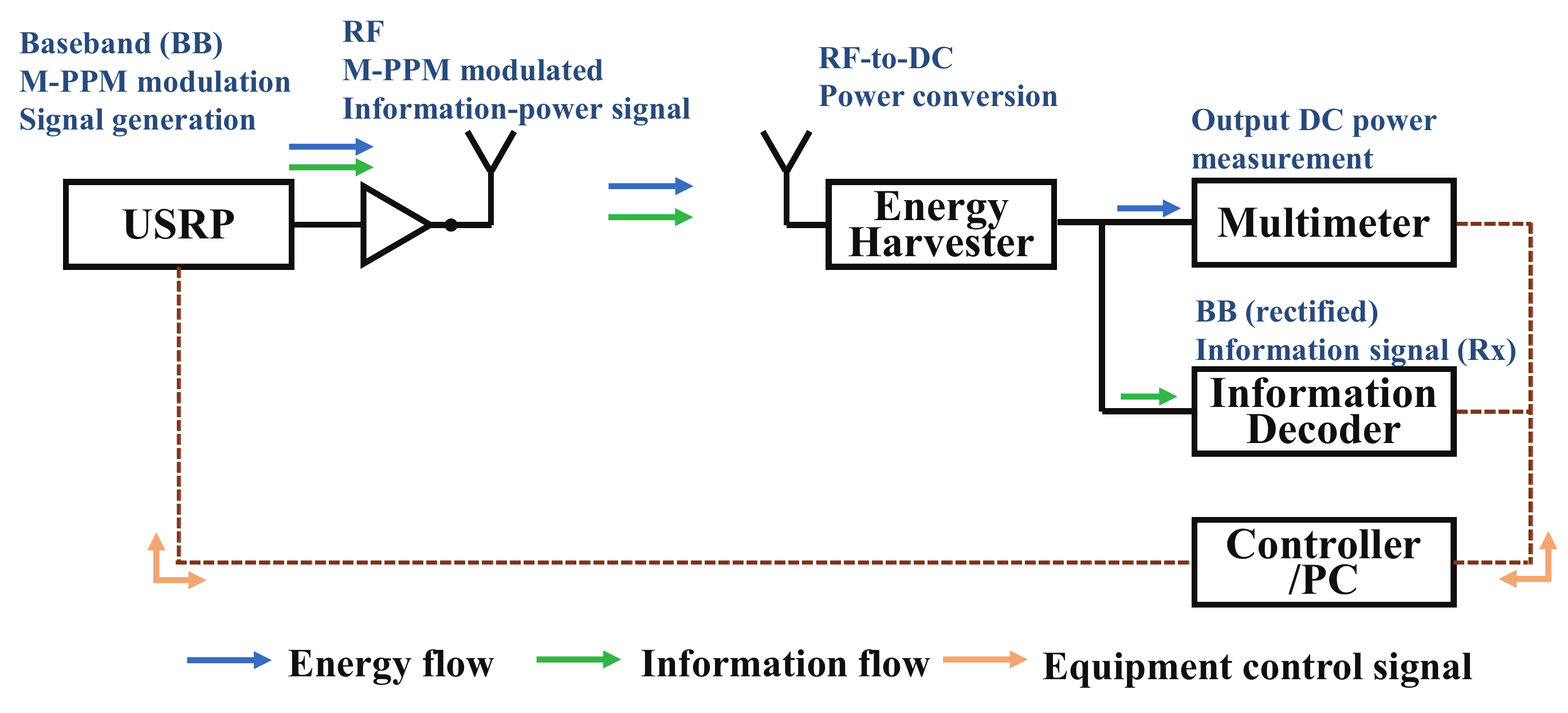}
	\caption{The prototype system configuration of the SWIPT with IntRx receiver (compatible with M-PPM modulation).}
	\label{fig_diagram}
\end{figure}

We implemented a point-to-point SWIPT (using IntRx receiver) prototype capable of using the newly proposed M-PPM modulation. 
A brief configuration of the SWIPT prototype system is illustrated in Fig.\ref{fig_diagram}, and the system consists of the transmitter, receiver, and system controller. 
The transmitter implemented using USRP is able to generate baseband and RF M-PPM modulated information-power signal and is also able to radiate the signal over-the-air. 
The receiver is the same in Fig. \ref{fig_systemmodel} and contains an EH to convert RF signals to DC and ID to demodulate the rectified M-PPM modulation signal. 
Additionally, the system controller is implemented to manage the system operation and to gather measurement results such as BER and harvested DC power.
We implemented and modified the system leveraging our existing WPT and SWIPT prototype reported in \cite{Kim2020} and \cite{Kim2019} to analyze whether the MPPM modulation is feasible in a realistic environment and to evaluate its performance for information and power transfer. 
\par
\begin{figure}[t]
	\centering
	\includegraphics[width=0.4\textwidth]{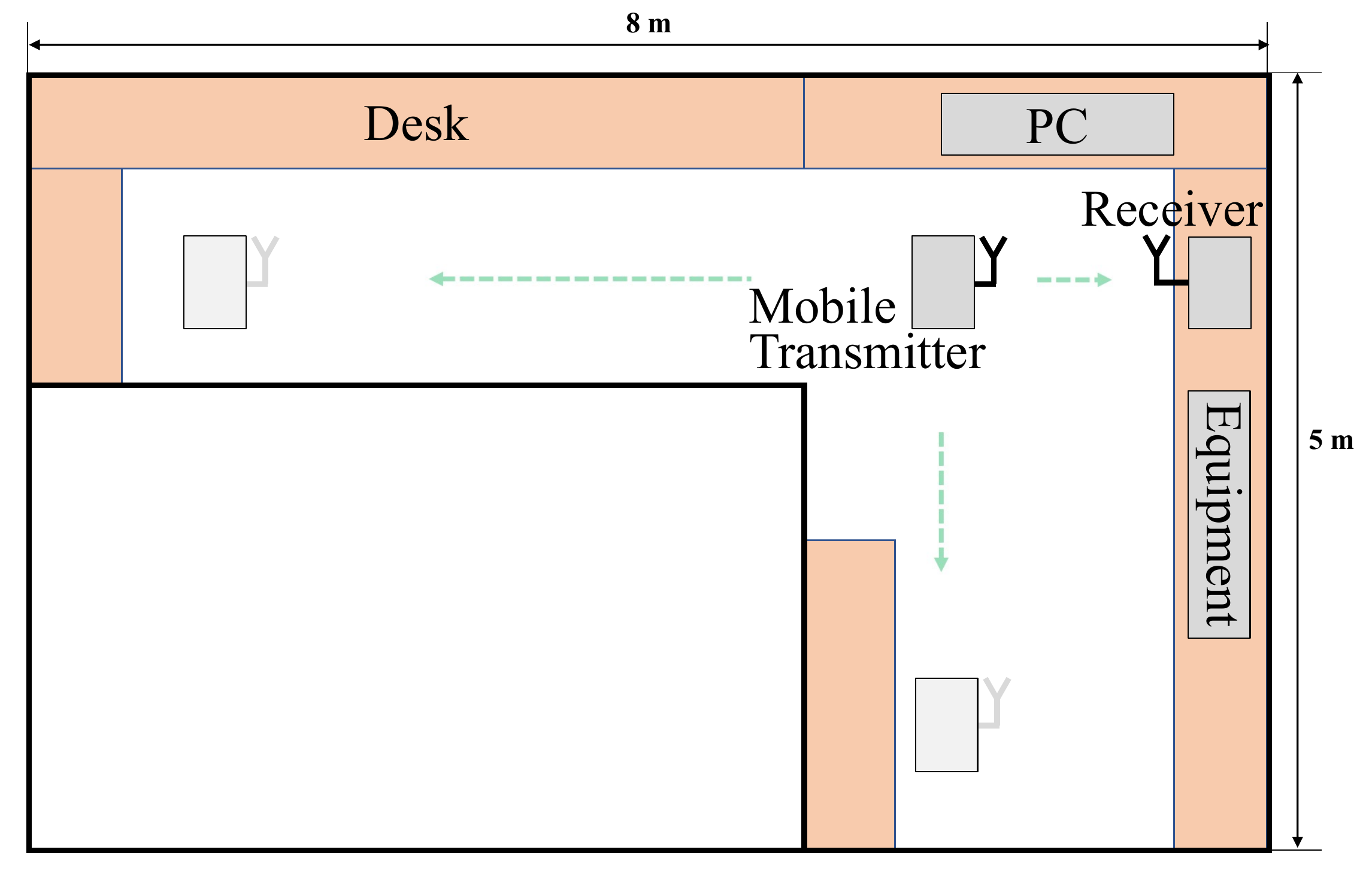}
	\caption{Illustration of the layout of SWIPT system prototype installed in an indoor office environment.}
	\label{fig_layout}
\end{figure}
\begin{figure}[t]
	\centering
	\includegraphics[width=0.48\textwidth]{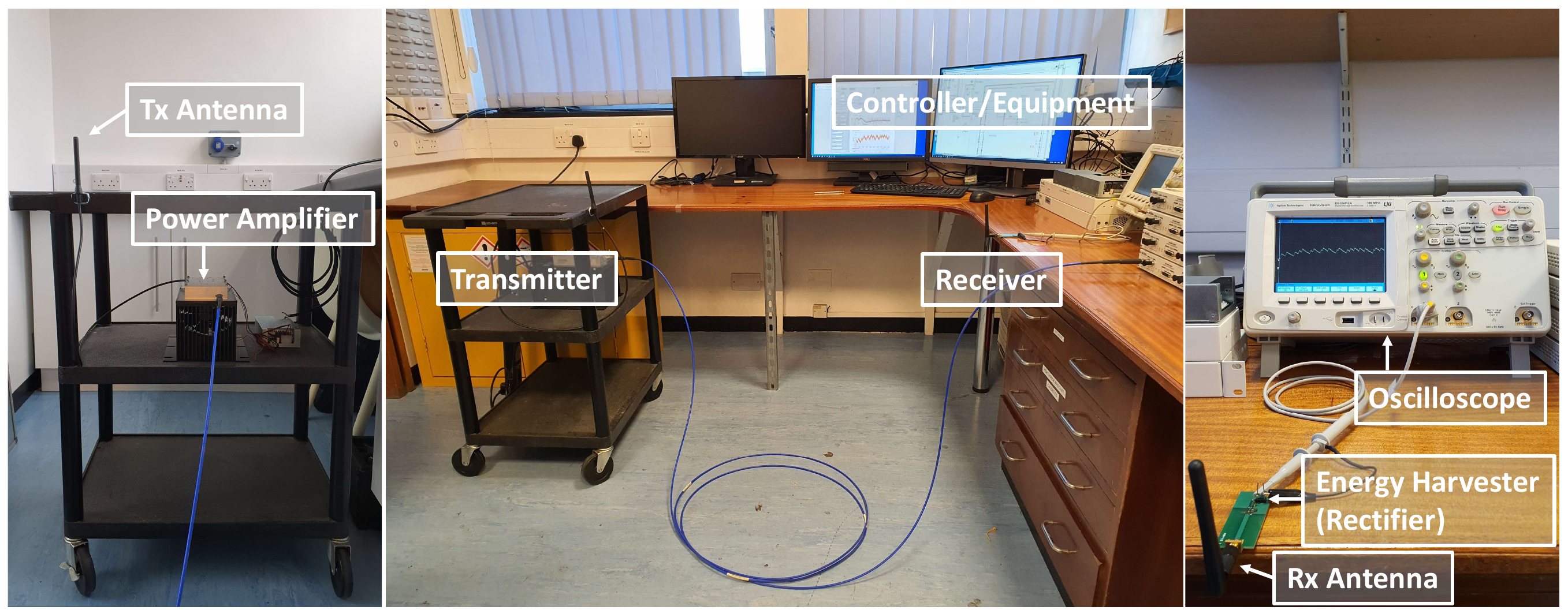}
	\caption{Photos of the SWIPT system prototype installed in an indoor office environment.}
	\label{fig_prototype}
\end{figure}

The SWIPT prototype system is installed in an 8m x 5m indoor laboratory with common facilities such as desks, chairs, PCs, and wardrobes that cause multipath fading in wireless channels.
The laboratory in which the prototype was installed is similar to a common indoor office environment, so it is very similar to the actual environment of a practical SWIPT system.
We adjoined mobility features to the system to enable performance evaluation on various wireless channels in the lab.
The location of the receiver is fixed, but the transmitting antenna is installed on a mobile trolley so that it can be located anywhere in the laboratory.
A brief layout of the indoor laboratory and photos of the installed system including the transmitter, receiver, and system controller are shown in Fig.\ref{fig_layout} and Fig.\ref{fig_prototype}, respectively.
\par
The system operates in the 2.4GHz ISM band.
The MPPM modulation and demodulation scheme have been applied to both transmitter and receiver. 
The transmitter is implemented using software-defined radio equipment USRP (NI USRP-2942), which is able to generate and radiate the M-PPM modulated SWIPT signal (various bandwidth and message size) as well as single-tone continuous wave and conventional modulation signals such as PSK and QAM. 
An external power amplifier (ZHL-16W-43+) is applied between the transmit antenna and the USRP to enhance the transmission power while complying with the maximum EIRP of 36dBm specified in FCC Title 47, Part 15 regulations \cite{FCC}.
The RF signal received by the receiving antenna is first input to the EH and converted into a DC voltage signal $v_{DC}(t)$.
The EH simply consists of a matching network, rectifying diode (Skyworks SMS7630), and low pass filter.
The values of the matching network and low pass filter circuit components were chosen to fit the 2.45GHz operation frequency, and the entire circuit is fabricated using 1.6mm thick FR-4 substrate and lumped elements.
We have used the same single-diode rectifier as the EH used in \cite{Kim2020}, reproduced in Fig. \ref {fig_prototype} with circuit components values of 0.3pF for $C1$, 2.4nH for $L1$, 10k$\Omega$ for $load$, and 1nF for $C_{out}$, and the EH with the smaller capacitor of 200pF for $C_{out}$ at the output low pass filter is also used for performance comparison similarly to what was conducted in the simulations for the previous section.
The rectified DC output signal is read by an oscilloscope for the demodulation process. 
The oscilloscope is an alternative to the ADC in the IntRx receiver structure presented in Fig. \ref{fig_systemmodel}.
To simplify the system implementation, the ADC and demodulator functions were replaced by using an oscilloscope and a PC instead of separate embedded equipment.
We used Agilent DSO5012A oscilloscope to read DC voltage signal and it supports up to 2GHz sampling rate. 
Digital input signal to ID $y[n]$ is read from the oscilloscope and is transferred to the PC to proceed with the demodulation process.
At the PC, the moving average filter is applied to the signal $y[n]$ and the information is demodulated, then the demodulated data bits are compared with the original bits to measure BER. 
Meanwhile, the harvested DC power is also calculated using the signal $y[n]$ which is the same as the measured voltage output from the EH. 
Entire measurement results can be recorded in real-time and stored. 
The operating sequence of the prototype system is implemented by LabView, and the entire system is controlled by a host PC.

\subsection{Information Transfer Performance Evaluation}

\par
\begin{figure}[t]
	\centering
	\includegraphics[width=0.4\textwidth]{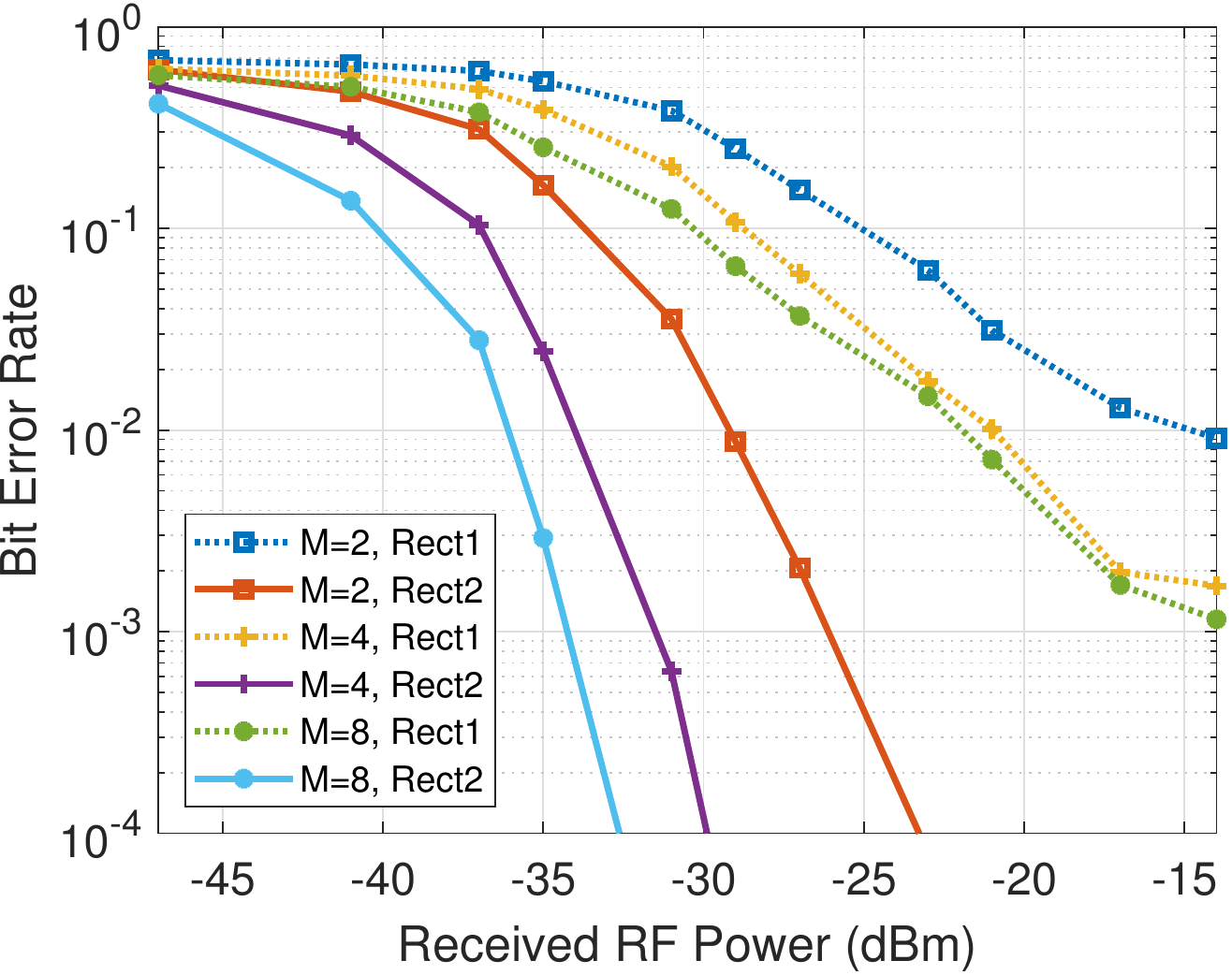}
	\caption{Measured BER performance versus received RF power of M-PPM signals with 5MHz bandwidth and three different modulation order (M = 2, 4, 8), Rect1 : $C_{out}$=1nF, Rect2 : $C_{out}$=200pF.}
	\label{fig_ppm_exp_ber_M}
\end{figure}

In this section, firstly, we have measured the BER performance of the M-PPM modulation signal while sweeping the input RF power level to the receiver to verify the information transfer performance experimentally. 
In order to accurately control the power of the received RF signals, the transmission and reception ports of the USRP equipment are connected using a coaxial cable with an attenuator, and the transmit power is digitally controlled by the host controller.
BER measurement results with three different modulation order ($M$ = 2, 4, and 8) of M-PPM modulation with a signal bandwidth of 5MHz and using two different output capacitor sizes of the EH are indicated in Fig. \ref{fig_ppm_exp_ber_M}. 
\par
The graph evidently demonstrates that the improved BER performance originates from the higher-order modulation and a smaller output capacitor of the rectifier. 
When the received signal power is around -20 dBm, the BER performance of M-PPM signals with a 1nF output capacitor in the EH is around $10^{-2}$. 
However, using a 200pF capacitor at the EH leads to significantly lower BER with the same signal design, and higher throughput can be achieved. 
Therefore, the throughput of 4-PPM modulation with 5MHz bandwidth using rectifier with 200pF output capacitor can be 2 Mbps in the realistic system with input RF power of -20dBm. Using a larger value of capacitor of 1nF leads to a smoother output signal and makes it sensitive to noise, which consequently leads to a drop in the data throughput to 1.996 Mbps.
These results are consistent with the simulation results displayed in the previous section. 

\par
\begin{figure}[t]
	\centering
	\includegraphics[width=0.4\textwidth]{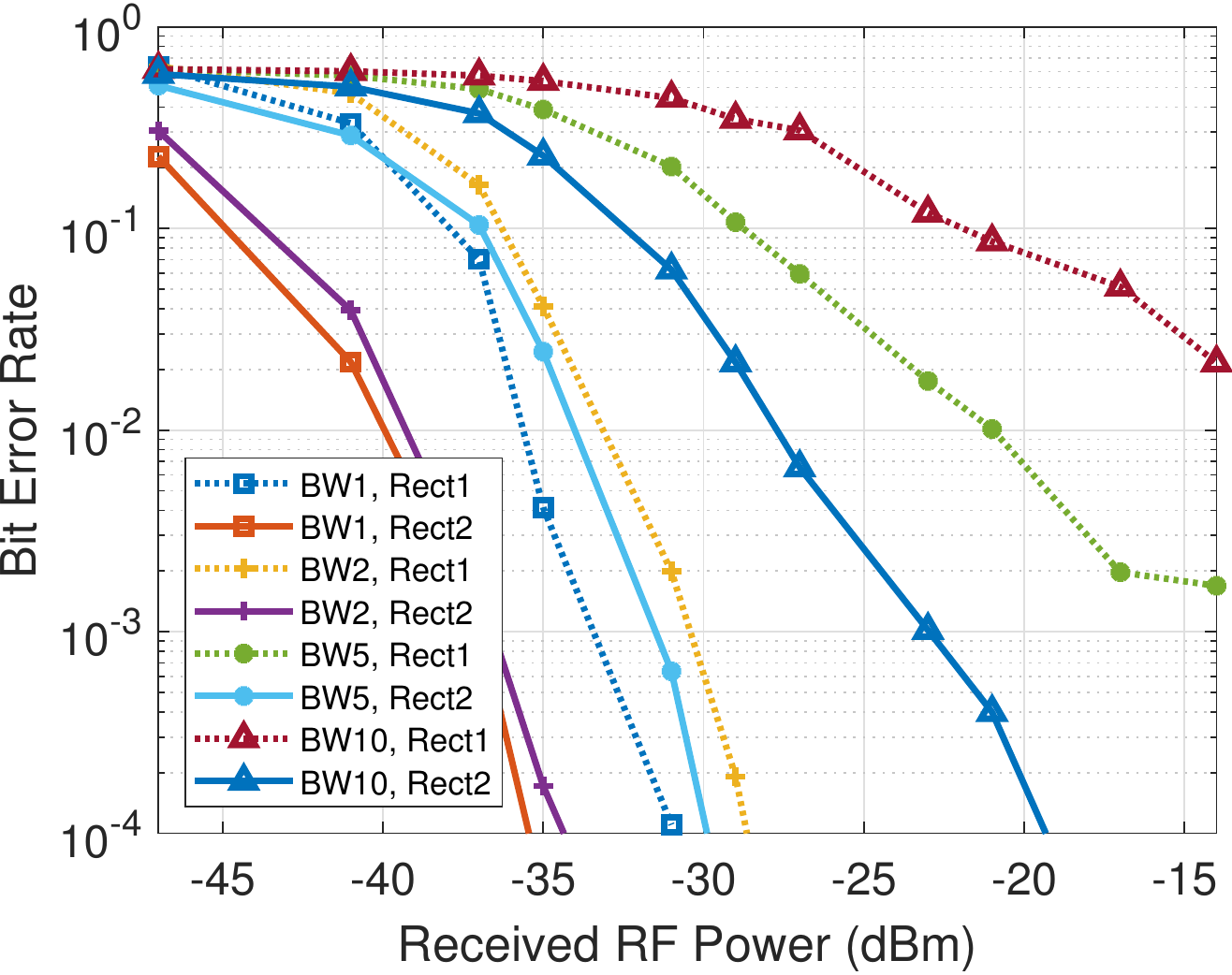}
	\caption{Measured BER performance versus received RF power of 4-PPM signals with different bandwidths, Rect1 : $C_{out}$=1nF, Rect2 : $C_{out}$=200pF.}
	\label{fig_ppm_exp_ber_BW}
\end{figure}

BER measurement results using different bandwidth signals of 4-PPM modulation and output capacitor sizes of EH are indicated in Fig. \ref{fig_ppm_exp_ber_BW}. 
The graph indicates that a better BER performance can be obtained with a smaller signal bandwidth and a smaller output capacitor of the EH, which is in line with the results confirmed by simulations. 
As discussed with the simulation results, the lower bandwidth signal has a longer period of pulse duration (charging capacitor at the EH) and a longer duration of the off transmission (discharging capacitor at the EH). 
This behavior magnifies the fundamental ripple of the signal at the output of the EH and makes the contained information more noticeable.
However, a reduced bandwidth reduces the data throughput.
This graph also clearly shows that the BER performance significantly increases when the output capacitor of the EH decreases for signals with the same bandwidth.
\begin{figure}[t]
	\centering
	\includegraphics[width=0.42\textwidth]{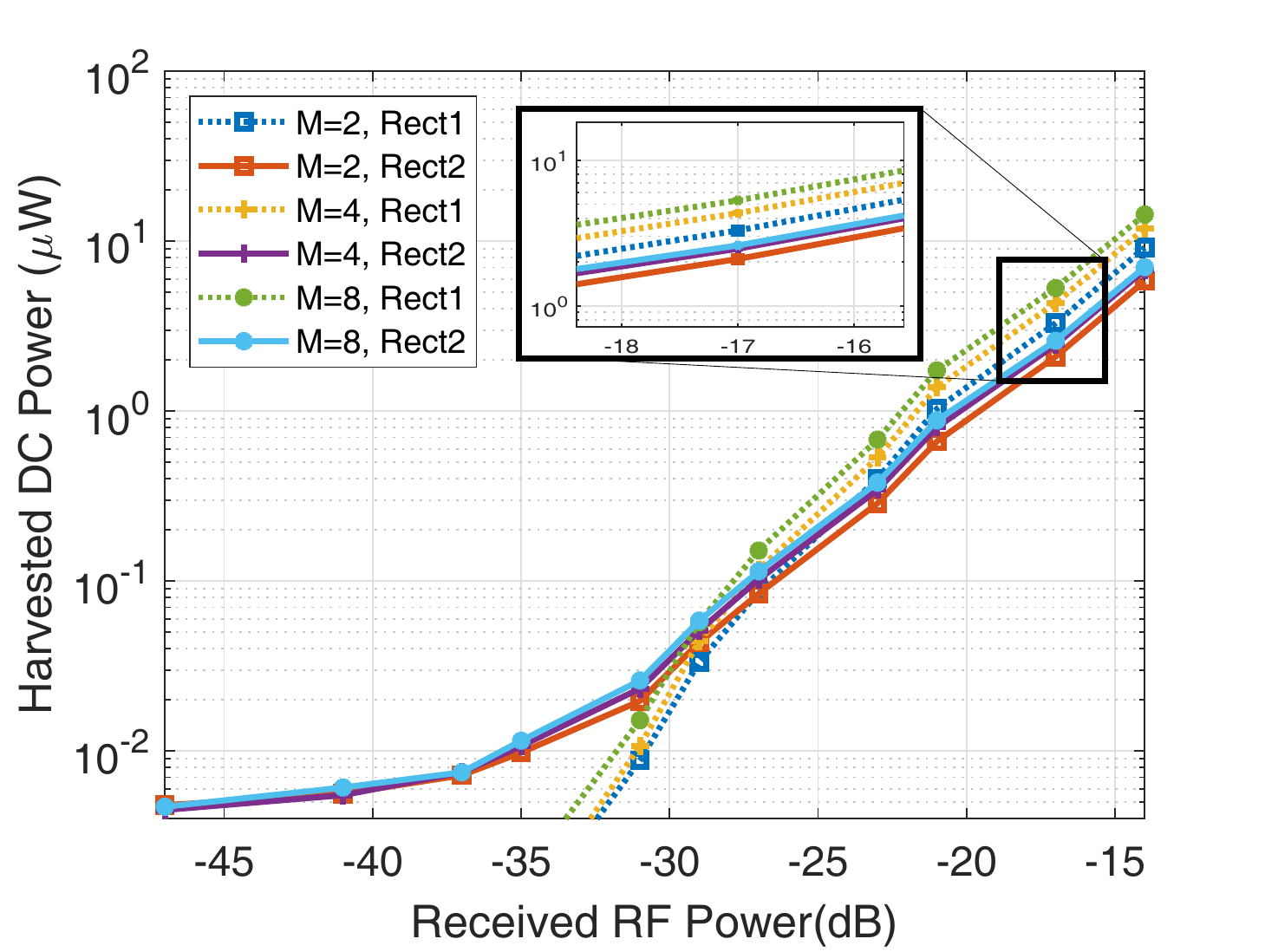}
	\caption{Measured harvested DC power versus received RF power of M-PPM signals with 5MHz bandwidth and three different modulation order (M = 2, 4, 8), Rect1 : $C_{out}$=1nF, Rect2 : $C_{out}$=200pF.}
	\label{fig_ppm_exp_pwr_M}
\end{figure}
\begin{figure}[t]
	\centering
	\includegraphics[width=0.42\textwidth]{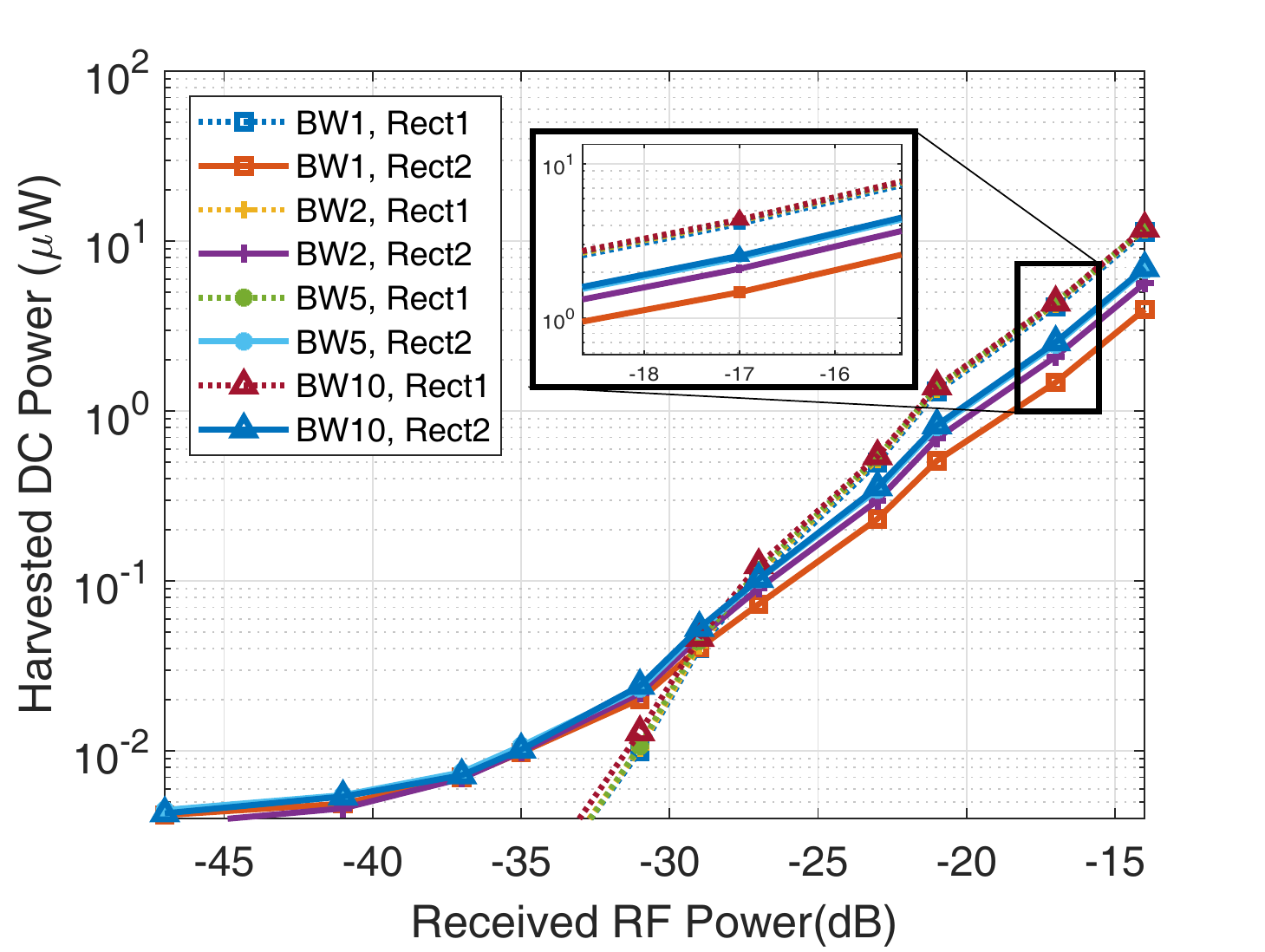}
	\caption{Measured harvested DC power versus received RF power of 4-PPM signals with different bandwidths, Rect1 : $C_{out}$=1nF, Rect2 : $C_{out}$=200pF.}
	\label{fig_ppm_exp_pwr_BW}
\end{figure}
\par
It is challenging to directly compare the information transfer performance of M-PPM signals for the SWIPT IntRx receiver to other conventional modulation schemes and RF receivers.
If compared with a simple form of BPSK modulation with conventional RF receiver, the BER performance at the same received RF power as well as the achievable information throughput is inevitably worse with M-PPM signals.
Conventional RF receivers and modulations can be operated even at a much lower received RF power level (less than -80dBm) and spectral efficiency of BPSK is higher with 1 bits/s/Hz rather than 0.4 bits/s/Hz of 4-PPM. 
This performance advantage of the conventional systems can be possible due to the various RF components such as LNA and mixer, and those RF components consume a few mW of power for their operation even in current commercial ultra-low-power RF transceivers.
Although the M-PPM modulation with IntRx SWIPT receiver is inferior to the conventional system in terms of information transfer performance, the huge benefits come from the significant reduction in power consumption of the information decoding module.
Low-power operation is the most emphasized design factor for the SWIPT receiver in many applications, so the benefits of M-PPM modulation with IntRx SWIPT receiver are worth highlighting.

\subsection{Power Transfer Performance Evaluation}

\par
In this section, we provide the harvested DC power measurement results and their analysis. 
The harvested DC power measurements were carried out with the BER performance measurements at the same time under the same condition.
The power measurement results with three different modulation order $M$ with 5MHz bandwidth and 4-PPM modulation with four different bandwidth $BW$ are displayed in Fig. \ref{fig_ppm_exp_pwr_M} and Fig. \ref{fig_ppm_exp_pwr_BW}, respectively. 
As expected from the theoretical and numerical analysis in the previous sections, Fig. \ref{fig_ppm_exp_pwr_M} confirms experimentally the WPT performance enhancement with a larger modulation order $M$.
The higher-order modulated signal outperforms the lower-order modulation in the range where the harvested energy is greater than 0.1 $\mu$W for both rectifiers.
Fig. \ref{fig_ppm_exp_pwr_BW} shows slightly different trends between the two rectifiers. 
The increase in bandwidth of the 4-PPM signal does not affect the WPT performance with rectifier one (with a large-sized capacitor $C_{out}$=1nF), whereas higher bandwidth signals clearly show a better WPT performance than lower bandwidth signals in rectifier two. 
Those performance variations with bandwidths between the two rectifiers are associated with the charging and discharging time at the rectifier's output capacitor.
Increasing the M-PPM signal's bandwidth shortens the symbol and pulse duration, i.e., the charging and discharging time of the rectifier output capacitor.
In rectifier one, the M-PPM signals with varying bandwidths scarcely affect the output waveform shape because the time-constant of the rectifier's low pass filter is sufficiently large to flatten the M-PPM input signals from 1MHz to 10MHz range.
However, in rectifier two, the output capacitor can be completely or nearly discharged during a symbol period of smaller bandwidth signals (i.e., longer symbol period) due to its small time-constant, and it causes degradation of DC output power performance. 
Recalling Fig. \ref{fig_fund_ripple}, the output voltage signal with 1 or 2 MHz bandwidth using the 200pF of $C_{out}$ significantly fluctuates, and its peak-to-peak fundamental ripple voltage is nearly identical to the peak-to-peak voltage of the signal. 
On the other hand, the signals with higher bandwidth of 5 and 10 MHz look much more flattened, similar to the larger $C_{out}$ EH case.
Therefore, the DC output power performance of a 4-PPM signal with 5 and 10 MHz bandwidth in rectifier 2 are nearly indistinguishable.
\begin{figure}[t]
	\centering
	\subfigure[]{\includegraphics[width=0.22\textwidth]{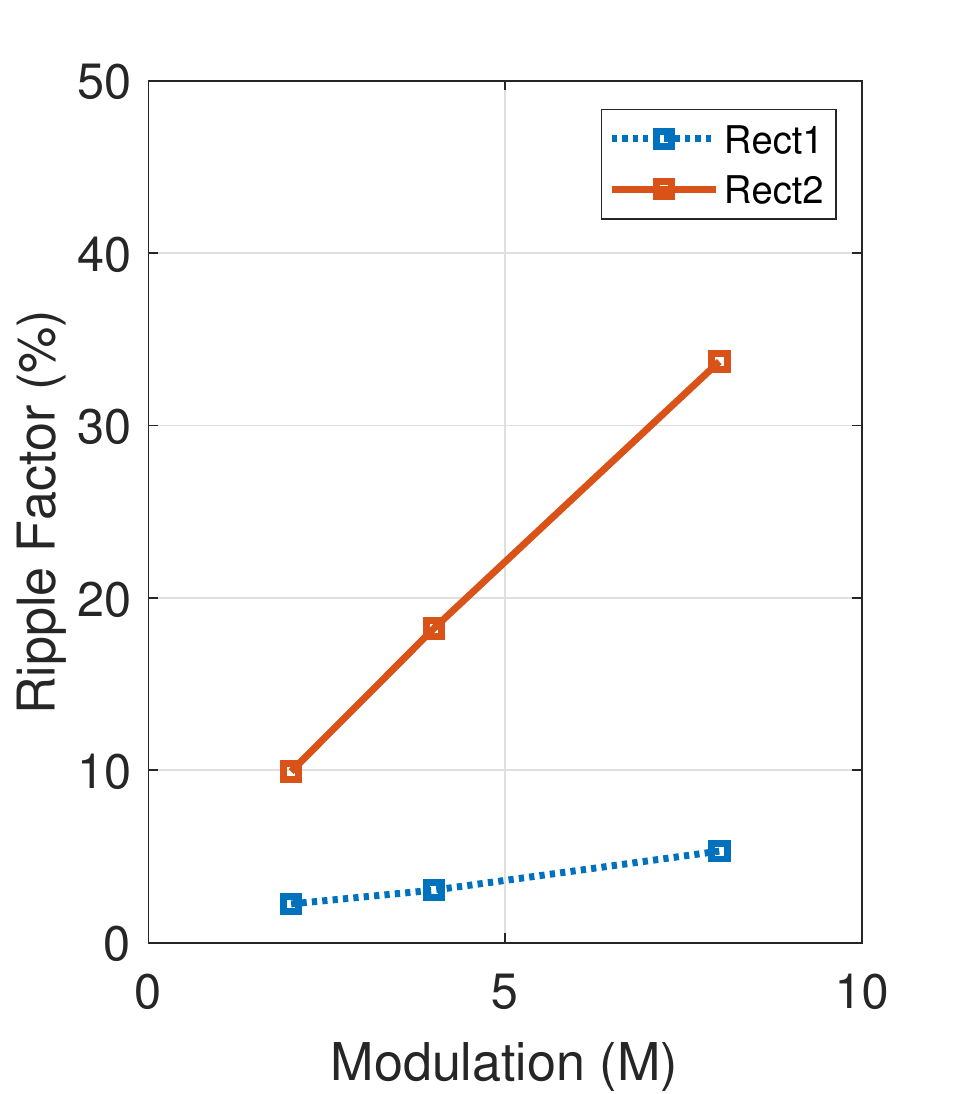}}
	\subfigure[]{\includegraphics[width=0.22\textwidth]{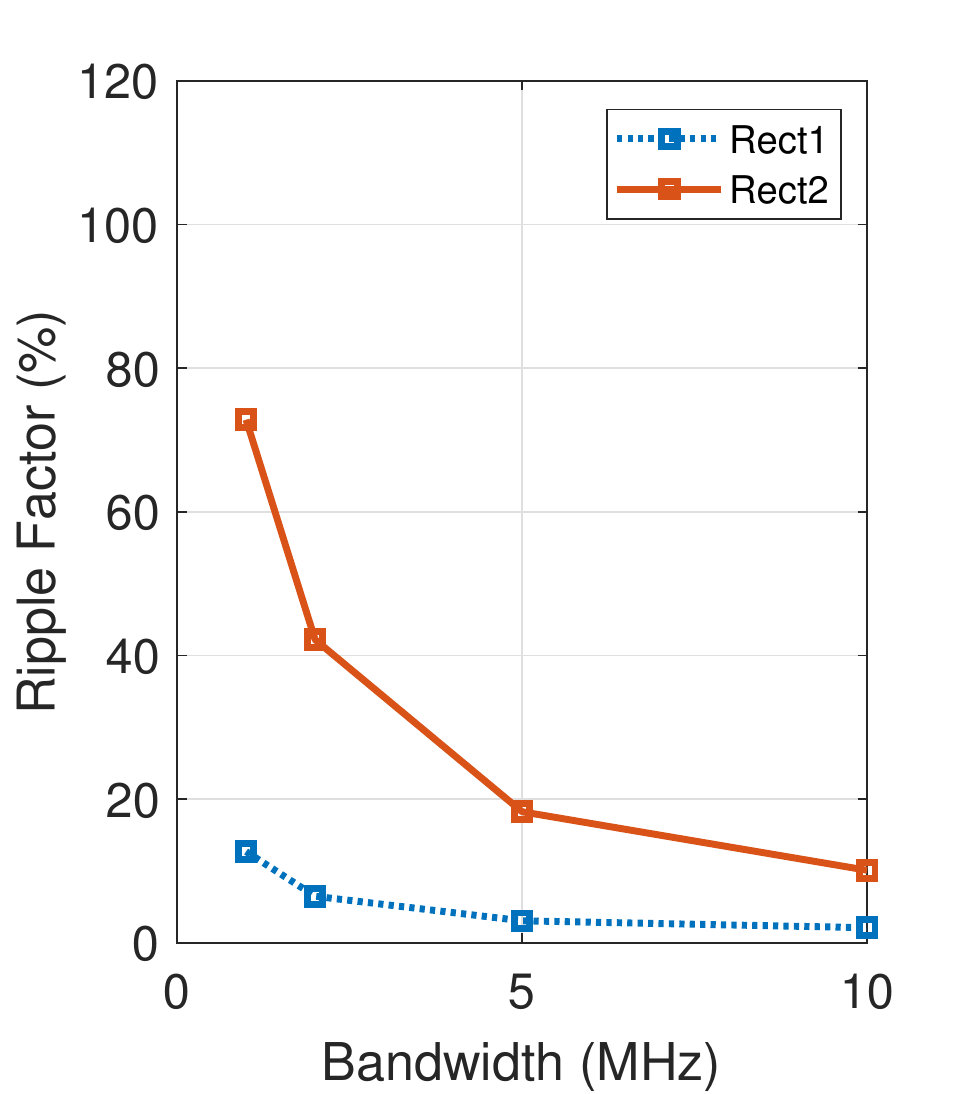}}
	\caption{Measured ripple factors of M-PPM modulated signals at -17dBm RF input power (a) versus modulation order (b) versus bandwidth, Rect1 : $C_{out}$=1nF, Rect2 : $C_{out}$=200pF.}
	\label{fig_exp_ripplefactor}
\end{figure}
\par
We also measured the ripple factors experimentally with the input power of -17 dBm signals with various bandwidths, modulation orders, and rectifier design.
Fig. \ref{fig_exp_ripplefactor} is a measurement result of the ripple factors, including fundamental ripple and noise ripple using the prototype. 
As theoretically and numerically confirmed earlier, the experimental measurement results in Fig. \ref{fig_exp_ripplefactor} also show the consistent trend that high ripple factors are obtained with a smaller bandwidth, a higher modulation order, and small rectifier output capacitors.
In order to supply stable DC power, the ripple factor of the harvested voltage signal should be maintained at a low level. 
Proper selections of signal bandwidth and rectifier characteristics are needed to realize the PPM signals as a viable power source for the practical SWIPT receiver. 
Therefore, we select the M-PPM signal with 5 MHz bandwidth and the rectifier with 1nF of $C_ {out}$ to experimentally evaluate the power transfer performance similar to what was done with the simulations in the previous section.
\begin{figure}[t]
	\centering
	\includegraphics[width=0.4\textwidth]{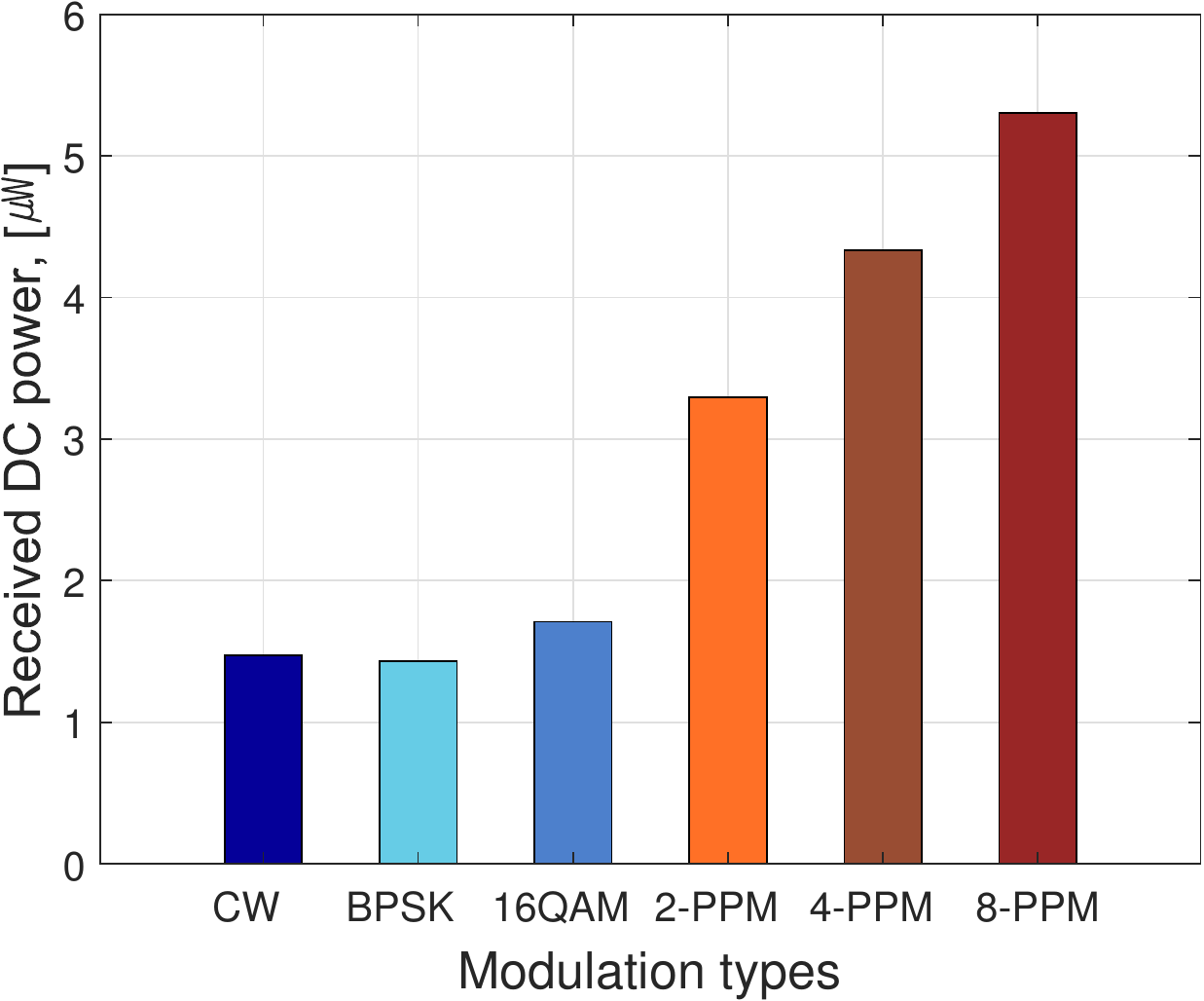}
	\caption{Harvested DC power performance of conventional and M-PPM modulation signals at -17dBm input RF power.}
	\label{fig_ppm_pwr_compare}
\end{figure}
Fig. \ref{fig_ppm_pwr_compare} displays the measured harvested DC power of several M-PPM signals and comparisons with conventional modulations such as PSK and QAM.
The measurement was carried out using 2, 4, 8-PPM signals with its input power of -17dBm, and the results clearly indicate the significant gains of M-PPM signals over CW and conventional modulations, in line with the theoretical analysis and numerical evaluation. 
The gain over CW is 123\%, 194\%, and 259\% for each 2, 4, 8-PPM signals, respectively.
The higher-order modulation shows a larger gain over CW.
As analyzed in the simulation results, the PAPR of the signal increases with the higher-order modulation since a larger amplitude pulse is transferred with a lower occurrence, and the output DC power increases thanks to the rectifier nonlinearity.
This result confirms that the enhancement in energy harvesting performance using M-PPM is remarkable and feasible in a realistic SWIPT system.

\par
Additional M-PPM's power transfer performance experiments were carried out under many different wireless environments.
We have measured output DC power at the receiver using M-PPM and conventional signals at 50 different locations in our testbed. 
The distance between the transmitter and the receiver varies between 0.5 and 5.5 meters, and the transmission power is fixed at 27dBm.
All six different types of signals were tested by sequentially repeating 50 times at a location without any changes of circumstance, before averaging the results for each signal.
So, at a specific location, the wireless channel status varied barely, and even if there is a change, all different types of test signals experience the same change. 
\begin{figure}[t]
	\centering
	\includegraphics[width=0.4\textwidth]{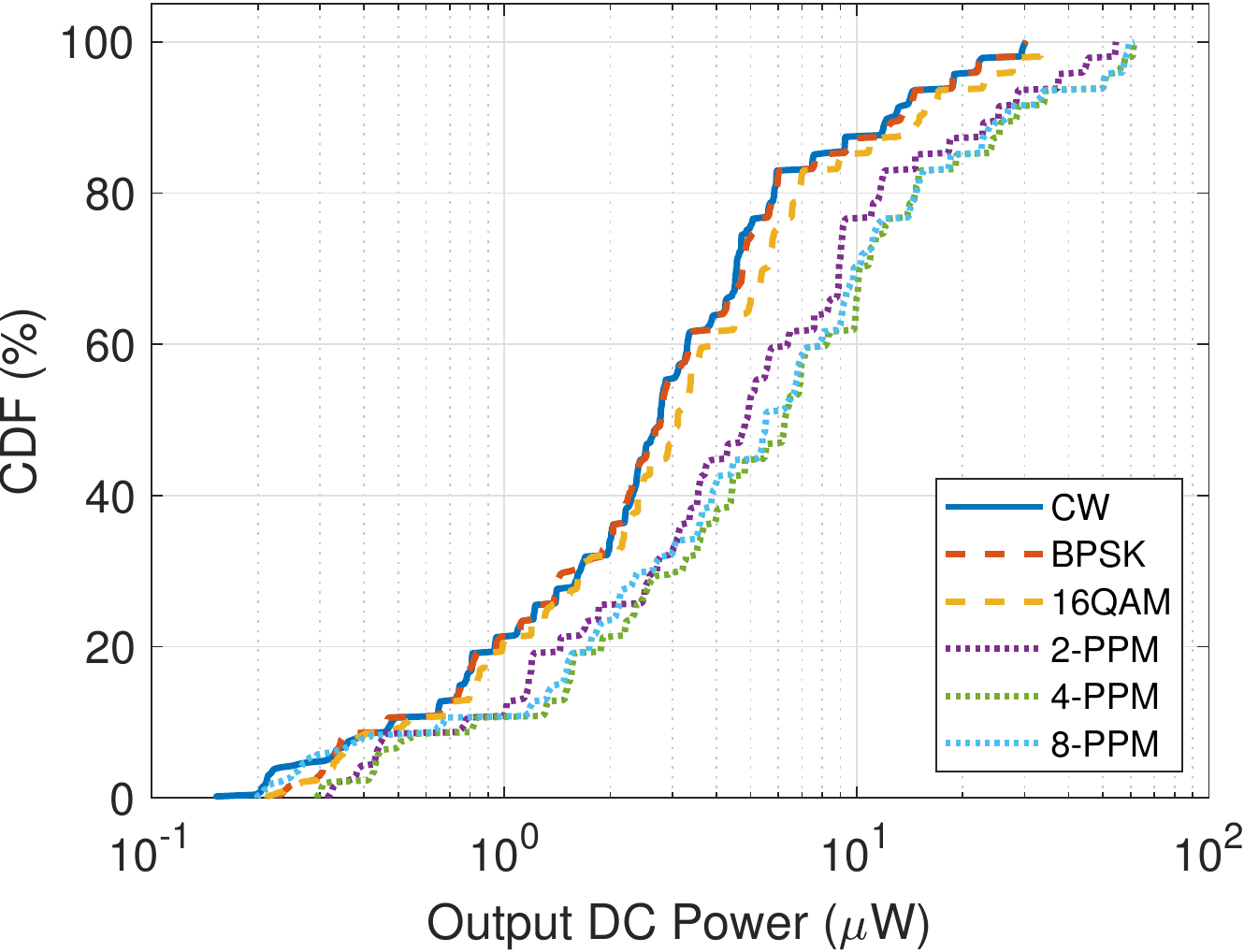}
	\caption{Measured CDF of output DC power at different locations in the indoor lab. environment.}
	\label{fig_ppm_pwr_cdf}
\end{figure}
Cumulative distribution functions (CDF) of measured DC output power with various modulation signals are presented in Fig. \ref{fig_ppm_pwr_cdf}. 
The graph clearly shows the rightly shifted CDFs of M-PPM signals compared to conventional ones, which means the measurement results of M-PPM are distributed in higher power regions.
Moreover, the higher-order M-PPMs are more shifted to the right.
Such behavior reaffirms the earlier observations we identified regarding the gain increase with higher-order modulation.
Those observations are inline with the simulation results in Fig. \ref{fig_ppm_sim_power} and the theoretical gain of M-PPM (mostly in the fourth-order term of the scaling law) according to Table \ref{table_scalinglaw}. 
This indicates that the theoretical analysis and simulation results are consistent with the experimental results in the actual wireless environment.

\section{Conclusion}\label{sec:conclusion}

We proposed a new type of modulation and decoding scheme called M-PPM for the SWIPT with IntRx architecture. 
The detailed structure of IntRx, which includes EH and ID, and its system models were presented, and the transmission signal generation method and information decoding scheme were described in detail and graphically explained. 
A realistic SWIPT testbed was established in an indoor environment using prototypes of a SWIPT transmitter and an IntRx receiver.
The performances of both power transfer and information transfer were evaluated by circuit simulations and also experiments using the testbed.
The harvested DC power and the information throughput achieved by M-PPM were analyzed as a function of various parameters such as the modulation orders, bandwidth, and rectenna design.
The experimental results were contrasted with the simulation and theoretical analysis.
This work has shown that the M-PPM is capable of boosting the harvested DC power performance as well as decoding information without power-consuming RF components, and it is feasible in real-world wireless conditions.
Large gains in WPT performance are obtained using M-PPM with a proper design of an energy harvester thanks to the rectifier nonlinearity, but there is also a trade-off between WPT and WIT performance. 
Various system parameters can be a design factor to meet and optimize the SWIPT system performance to the various requirements, or it can be adaptively chosen for multiple purposes.
This work comprehensively demonstrates the novel signal strategy M-PPM for the IntRx SWIPT system from theory to experiment and draws a consistent performance analysis from those different methodologies.
In other words, by confirming that the theoretical modeling and simulations match the experimental results in a realistic wireless environment, the usefulness of this new form of signal design for SWIPT system is demonstrated, and the nonlinear rectenna model proposed in previous studies is once again verified.

\par
There are many interesting future works to be investigated, including the design and implementation of M-PPM for multi-antenna SWIPT settings, CSI adaptive M-PPM, M-PPM receivers with off-the-shelf components such as ADC, MCU, and sensors, as well as the application of M-PPM-based SWIPT in sensor networks and IoT solutions. 

\bibliographystyle{IEEEtran}
\bibliography{jhlib}

%

\end{document}